\documentclass{article}
\usepackage{PRIMEarxiv}
\usepackage[utf8]{inputenc} 
\usepackage{newunicodechar}
\pdfoutput=1
\usepackage[T1]{fontenc}    
\usepackage{hyperref}       
\usepackage{url}            
\usepackage{booktabs}       
\usepackage{amsfonts}       
\usepackage{nicefrac}       
\usepackage{microtype}      
\usepackage{lipsum}
\usepackage{fancyhdr}       
\usepackage{graphicx}       
\graphicspath{{media/}}     

\title{Organic Electronics in Biosensing: A Promising Frontier for Medical and Environmental Applications
}

\author{
  Jyoti Bala Kaushal, Pratima Raut \\
  Department of Biochemistry and Molecular Biology,\\
  University of Nebraska Medical Center, \\
   Omaha, NE, 68198, USA\\
   \And
  Sanjay Kumar \\
  Durham School of Architectural Engineering and Construction,  \\
  University of Nebraska-Lincoln,  \\
  Scott Campus, Omaha, NE 68182-0816, USA\\
  \texttt{sanjay21505@gmail.com; skumar13@unl.edu} \\
}

\begin{document}
\maketitle

\begin{abstract}
The promising field of organic electronics has ushered in a new era of biosensing technology, offering a promising frontier for applications in both medical diagnostics and environmental monitoring. This review paper provides a comprehensive overview of organic electronics' remarkable progress and potential in biosensing applications. It explores the multifaceted aspects of organic materials and devices, highlighting their unique advantages, such as flexibility, biocompatibility, and low-cost fabrication. The paper delves into the diverse range of biosensors enabled by organic electronics, including electrochemical, optical, piezoelectric, and thermo sensors, showcasing their versatility in detecting biomolecules, pathogens, and environmental pollutants. Furthermore, integrating organic biosensors into wearable devices and the Internet of Things (IoT) ecosystem is discussed, offering real-time, remote, and personalized monitoring solutions. The review also addresses the current challenges and future prospects of organic biosensing, emphasizing the potential for breakthroughs in personalized medicine, environmental sustainability, and the advancement of human health and well-being.
\end{abstract}


\section{Introduction}
Organic electronics have emerged as a promising frontier in the field of biosensing, offering innovative and versatile solutions for medical and environmental applications. With the rapid advancement of organic materials and devices, integrating organic electronics into biosensing platforms has unlocked many possibilities for sensitive, real-time, and label-free biological and chemical analytes detection. This convergence of organic electronics and biosensing can revolutionize medical diagnostics, point-of-care testing, wearable health monitoring, and environmental monitoring, among other critical domains.

Based on carbon-based compounds and polymers, organic electronic devices present distinct advantages that make them well-suited for biosensing applications. These materials offer biocompatibility, enabling direct interactions with biological systems without causing adverse reactions, making them ideal for implantable biosensors and in-vivo monitoring. Additionally, organic materials exhibit exceptional flexibility, enabling the development of conformable and wearable biosensing devices that can seamlessly adapt to the human body or environmental surfaces, expanding their utility in personalized healthcare and environmental monitoring. The unique electronic properties of organic materials, such as tunability, conductivity, and semiconducting behavior, contribute to their exceptional sensing capabilities. Organic electronic devices, such as organic field-effect transistors (OFETs), organic electrochemical transistors (OECTs), and organic photodetectors (OPDs), have demonstrated high sensitivity, selectivity, and rapid response times, allowing for the accurate detection of target analytes in complex samples.

In this context, this review explores the exciting frontier of organic electronics in biosensing, focusing on its applications in medical diagnostics and environmental monitoring. We delve into an overview of organic bioelectronic materials, the various organic electronic devices, and their fabrication methods, detailing their sensing mechanisms and advantages over traditional sensing technologies. Additionally, we discuss the challenges faced in integrating organic electronics in biosensing platforms, such as biocompatibility, stability, manufacturing scalability, and data security and privacy, and the innovative strategies employed to address these obstacles.

\section{Organic Bioelectronic Materials}

\subsection{Conducting Polymers}
Conducting polymers (CPs) are a class of organic materials that exhibit electrical conductivity while maintaining the desirable mechanical properties of polymers. Unlike traditional semiconductors, CPs are intrinsically conductive without requiring any additional dopants. This unique combination of electrical and mechanical properties makes conducting polymers highly attractive for various applications, including electronics, biosensors, actuators, and energy storage devices. The electrical conductivity of conducting polymers arises from the delocalization of $\pi$ electrons, which occurs through the presence of alternating single and double bonds along their polymer chains. These $\pi$ electrons can move freely through the conjugated system, allowing the movement of charge carriers (electrons and holes) consequently resulting in electrical conductivity. CPs possess a valence band (HOMO - Highest Occupied Molecular Orbital) and a conduction band (LUMO - Lowest Unoccupied Molecular Orbital) \cite{le2017electrical}. The energy gap between the HOMO and LUMO determines the material's bandgap, affecting its electrical properties \cite{bredas2002organic}. In their pure, undoped state, organic polymers may behave as insulators or semiconductors due to the large energy gap between the HOMO and LUMO \cite{park2015review}.  Nonetheless, doping allows these materials to become conductive. Doping entails the introduction of additional charge carriers into the material, achieved through the incorporation of electron donors (n-type doping) or electron acceptors (p-type doping). This deliberate addition of charge carriers reduces the energy gap, thereby facilitating the movement of charge carriers and enhancing the electrical conductivity of organic polymers. Importantly, it's worth noting that not all organic semiconductors necessitate doping to exhibit their desired electrical properties.

In the case of CPs, the energy gap (bandgap) between the HOMO and LUMO is relatively small compared to insulators but larger than true metals. CPs exhibit distinct electrical and mechanical properties, allowing researchers to tailor their performance for specific applications. The electrical conductivity of conducting polymers can be tuned by varying factors such as oxidation state, doping level, and environmental conditions. CPs can be chemically doped or electrochemically doped to enhance their conductivity. By doping, additional charge carriers are introduced into the material, increasing electrical conductivity. Moreover, the mechanical properties of CPs are influenced by factors such as molecular weight, chemical structure, and processing methods. These polymers can be synthesized into various forms, including films, fibers, and coatings, while retaining conductivity. The flexibility and ease of processability make conducting polymers suitable for applications where traditional inorganic conductors may be limited due to their rigidity. The unique combination of electrical conductivity and mechanical flexibility enables conducting polymers to be used in electronic devices such as organic transistors, flexible displays, and printed circuits. They are also employed as sensing elements in chemical and biosensors, where their conductivity changes upon interaction with specific analytes. In the field of energy storage, conducting polymers are explored for applications in supercapacitors and batteries due to their high charge storage capacity.

\begin{figure}
\centering
\includegraphics[width=0.8\textwidth]{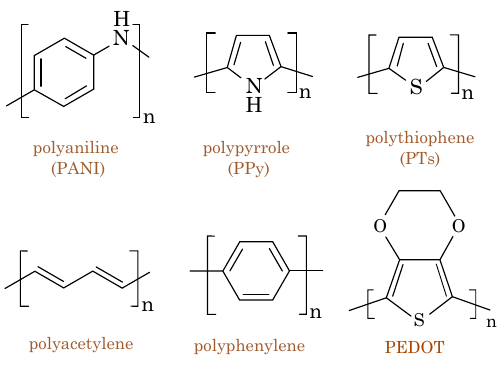}
\caption{Chemical structure of commonly used conducting polymers: polyaniline (PANI), polypyrrole (PPy), polythiophene(PTs), polyacetylene, polyphenylene, and poly(3,4-ethylenedioxythiophene) (PEDOT).}
\label{fig1}
\end{figure}

One of the pioneering conducting polymers is polyaniline (PANI), which first discovered its conductive properties in the late 1970s. Since then, several other conducting polymers, such as polythiophene(PTs), polypyrrole (PPy), and poly(3,4-ethylenedioxythiophene) (PEDOT), have been developed and extensively studied \cite{ramanavivcius2006electrochemical}.  PEDOT is the most ubiquitous organic mixed ionic/electronic conductors (OMIECs), a class of materials exhibiting simultaneous electronic and ionic conductivity \cite{paulsen2020organic}. This unique combination of properties makes OMIECs highly valuable for various applications, including electrochemical devices, energy storage systems, actuators and artificial muscles, and biosensors \cite{lu2023organic}. OMIECs comprise soft organic materials, such as conducting polymers or small organic molecules, that can conduct electrons and ions, offering advantages over traditional electronic or ionic conductors \cite{wu2022operando}. Figure~\ref{fig1} shows the chemical structure of commonly used conducting polymers.

\subsection{Organic Semiconductors}
Organic semiconductors are a class of organic materials with unique electronic properties, lying between traditional conductors and insulators. The band gap between organic semiconductors' valence band and conduction band is relatively lower than in insulators and higher than in conducting polymers. The organic semiconductors are composed of carbon-based molecules or polymers, known as $\pi$-conjugated systems, which enable the movement of charge carriers (electrons and holes) through their conjugated molecular structure \cite{brutting2005physics,kohler2015electronic,kunkel2021active,chen2023recent}. 

Small-molecule semiconductors consist of discrete, well-defined organic molecules, while polymer-based semiconductors comprise long-chain polymer structures with repeating monomer units. Organic molecules primarily comprise carbon atoms bonded to hydrogen, oxygen, nitrogen, and other elements. Carbon's ability to form stable covalent bonds with various other atoms allows for the diverse and complex structures found in organic molecules. Organic molecules contain specific functional groups, which are arrangements of atoms that confer distinct chemical properties and reactivity to the molecule. For example, the hydroxyl group (-OH) in alcohols is hydrophilic, while the carbonyl group (-C=O) in ketones and aldehydes has unique reactivity. 

In recent years, several organic molecular semiconductors have been extensively studied, including oligoacenes, oligothiophenes, discotic liquid crystals, triphenylamines, perylenes, tetrathiafulvalenes, and fullerenes \cite{mishra2012small}. Similarly, prominent examples of organic polymeric semiconductors include polyparaphenylenevinylene (PPV), polyparaphenylene (PPP), polyfluorene (PF), and polyfluorene copolymers \cite{coropceanu2007charge}. Both organic semiconductors offer advantages, such as solution processability and low-temperature deposition, making them suitable for various electronic and optoelectronic applications. Their versatile properties contribute to their widespread use in developing innovative and cost-effective devices for modern technologies.

\begin{figure}
\centering
\includegraphics[width=0.9\textwidth]{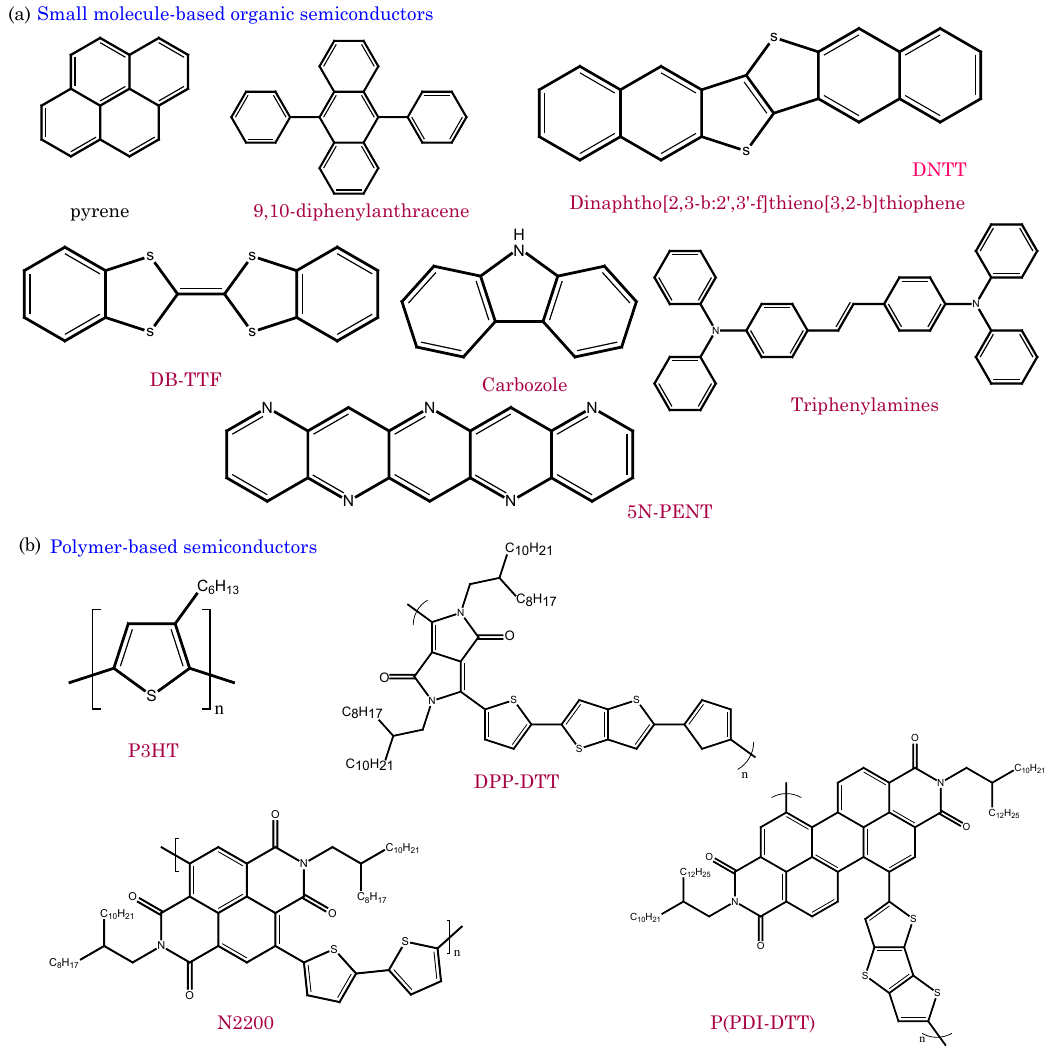}
\caption{Chemical structures of organic semiconductors \textbf{(a)} $\pi$-conjugated small molecular families based semiconductors, and \textbf{(b)} polymer-based semiconductors.}
\label{fig2}
\end{figure}

The unique properties of organic semiconductors have led to their integration into a wide range of electronic devices, such as organic field-effect transistors (OFETs) \cite{sun2005advances,zhang2022organic}, organic light-emitting diodes (OLEDs) \cite{dodabalapur1997organic,gather2011white,song2020organic}, organic photovoltaics (OPVs) \cite{brabec2004organic,kippelen2009organic,inganas2018organic}, and organic sensors \cite{wu2019highly}. OFETs utilize organic semiconductors as the active channel material, enabling flexible and low-power transistor devices. Organic sensors utilize the sensitivity of organic semiconductors to detect changes in environmental parameters, such as gas concentration or biomolecular interactions \cite{nasri2021gas}. Figure~\ref{fig2} shows examples of commonly used small molecule-based and polymer-based organic semiconductors for different types of bioelectronics devices \cite{hopkins2019all,borges2019organic,ta2022organic}.

While organic semiconductors have numerous advantages, such as flexibility and cost-effectiveness, they are not without challenges. These challenges encompass relatively lower charge carrier mobility when compared to their inorganic counterparts and susceptibility to environmental factors like humidity and temperature, as noted in recent studies \cite{kim2022new,subbarao2016organic,za2017effect}. To overcome these limitations, researchers are actively exploring advanced material engineering, innovative doping techniques, and novel device architectures to enhance the performance and stability of organic semiconductors \cite{kimpel2021conjugated,dong2023anion}.

Organic semiconductors remain a highly promising platform for developing flexible, cost-effective, and energy-efficient electronic devices \cite{neupane20192d}. Their unique properties and versatile applications have positioned them as compelling candidates for the next generation of electronic and optoelectronic innovations. These advancements drive innovation across various domains, including wearable technology, flexible displays, and renewable energy solutions. As research and development in organic semiconductors continue to progress, new opportunities are anticipated to emerge in the ever-evolving realm of organic electronics.

\subsection{Biomolecules as Sensing Elements}
Biomolecules serve as highly sensitive and selective sensing elements in various biosensing applications. These natural macromolecules, including proteins, nucleic acids, enzymes, and antibodies, exhibit specific interactions with target analytes, enabling the detection and quantification of various substances with remarkable accuracy. The inherent recognition capabilities of biomolecules make them valuable sensing elements in biosensors, enabling real-time monitoring of biochemical reactions and detecting analytes with exceptional specificity.

One of the key advantages of using biomolecules as sensing elements is their ability to bind specifically to target molecules, known as ligands or antigens, through molecular recognition processes \cite{ranallo2023synthetic}. This binding interaction is governed by complementary shapes and chemical properties between the biomolecule's active sites and the target analyte, allowing for highly selective detection \cite{kumar2018tapered}. The high affinity of biomolecules to their target analytes ensures that biosensors can distinguish between similar molecules, achieving precise and reliable measurements. Various techniques are employed to immobilize biomolecules onto the sensor surface while maintaining biological activity. Surface modification methods, such as physical adsorption, covalent binding, and self-assembled monolayers, allow the biomolecules to remain functional while attached to the sensor surface \cite{chaki2002self,sonawane2016surface,sandhyarani2019surface,li2020recent}. Immobilization ensures the biomolecular sensing elements remain near the transducer, facilitating efficient signal transduction upon analyte binding.

Moreover, enzymes are a specific class of biomolecules extensively used in biosensing applications due to their catalytic activity \cite{rocchitta2016enzyme,gonzalez2022biosensor}. Enzymatic biosensors utilize enzymes as sensing elements with a transducer to generate a detectable signal proportional to the concentration of the target analyte. This resulting signal arises from enzymatic reactions that induce changes in proton concentration, gas release/uptake (e.g., ammonia or oxygen), light emission, heat release, and more \cite{mulchandani1998principles}. The transducer converts this signal into a measurable response, like current, potential, temperature change, or light absorption, using electrochemical, thermal, or optical methods. This signal can be amplified, processed, or stored for subsequent analysis.

Additionally, antibodies are highly specific recognition elements used in immunoassays \cite{zeng2012recombinant,sharma2016antibodies}. They can selectively bind to antigens, pathogens, toxins, or specific biomolecules, forming antibody-antigen complexes. These complexes are detectable through various transduction methods, such as optical, electrochemical, or piezoelectric signals, allowing for sensitive and specific detection of the target analyte.

\begin{figure}
\centering
\includegraphics[width=0.9\textwidth]{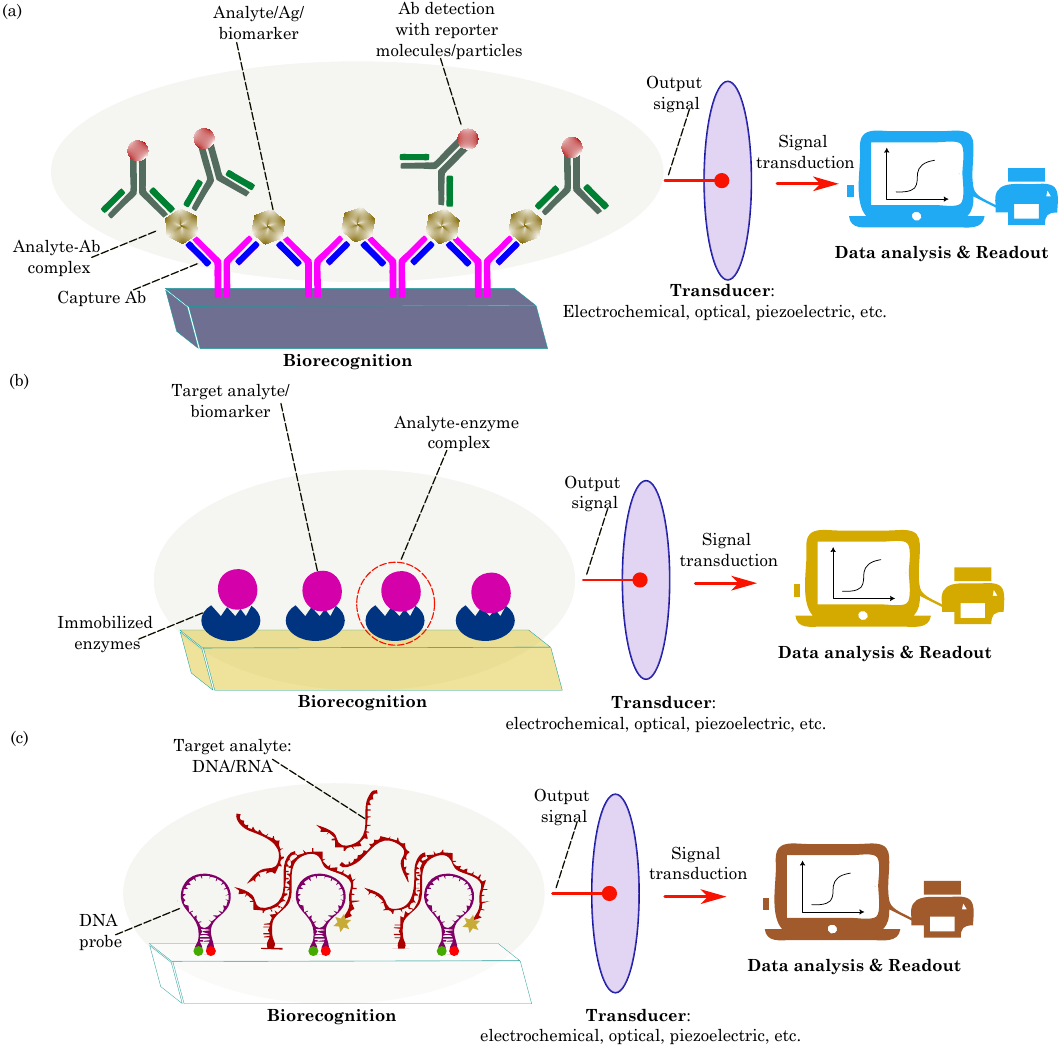}
\caption{Schematics illustration of biomolecules-based biosensors. \textbf{(a)} 
antibody-based; \textbf{(b)} enzyme-based biosensors; \textbf{(c)} DNA/RNA-based biosensors.}
\label{fig3}
\end{figure}

Furthermore, nucleic acids, such as DNA and RNA, are utilized in nucleic acid-based biosensors \cite{hahn2005nucleic,palchetti2008nucleic,du2017nucleic,fu2019recent}. These sensing elements recognize specific DNA sequences or RNA targets through hybridization reactions. Nucleic acid biosensors are vital for genetic analysis, disease diagnostics, and monitoring of nucleic acid-based biomarkers. Figure~\ref{fig3} illustrates the biomolecules-based biosensors. 

Overall, biomolecules serve as powerful sensing elements in biosensors, enabling the detection of a wide range of analytes, including proteins, nucleic acids, small molecules, and even viruses or bacteria. Their high specificity, sensitivity, and ability to function under physiological conditions make them invaluable tools in medical diagnostics, environmental monitoring, food safety, and various other applications. As research continues, integrating biomolecules into novel sensing platforms promises to revolutionize biosensing technology, opening up new avenues for precise, rapid, and cost-effective detection of analytes in diverse fields.

\subsection{Nanomaterials}
Nanomaterials are materials characterized by nanoscale dimensions, typically ranging from 1 to 100 nanometers in at least one dimension \cite{kumar2018fabrication}. These materials exhibit unique properties that differ significantly from their bulk counterparts, making them valuable for various science, engineering, and technology applications. The small size of nanomaterials results in a high surface-to-volume ratio, leading to enhanced reactivity and increased surface area for interactions with other materials. This unique feature allows for tailoring their physical, chemical, and mechanical properties through precise size, shape, and composition control \cite{ramya2022recent}.

\begin{figure}
\centering
\includegraphics[width=0.95\textwidth]{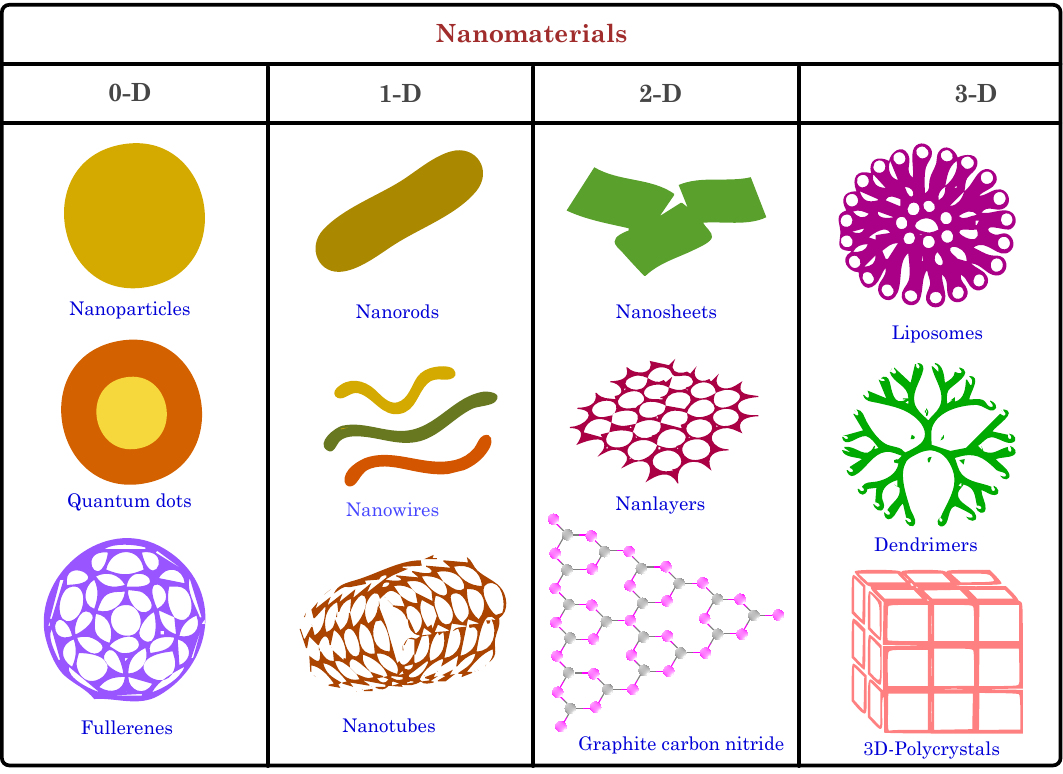}
\caption{Schematic illustration of nano-structured materials classified based on dimensionality.}
\label{fig4}
\end{figure}

Based on dimensionality, nanomaterials can be categorized into four main categories: zero-dimensional (0D), one-dimensional (1D), two-dimensional (2D), and three-dimensional (3D) nanomaterials (see Figure~\ref{fig4}). Zero-dimensional nanomaterials are nanoparticles with nanoscale dimensions in all three directions. Examples include nanoparticles and quantum dots. Nanoparticles comprise metals, metal oxides, semiconductors, polymers, and other materials. Due to their small size, nanoparticles exhibit quantum confinement effects, where their electronic and optical properties become size-dependent. This phenomenon leads to novel optical, electrical, and magnetic behaviors different from bulk materials. For example, gold nanoparticles exhibit unique plasmonic properties, making them suitable for applications in sensing and imaging \cite{jain2007review}. Another example of 0D nanomaterials is nanocomposites, formed by combining nanoparticles with a matrix material to enhance specific properties. These materials integrate the unique properties of nanoparticles, such as enhanced surface area and tailored functionality, with the structural support of the matrix material \cite{liu2020high,stephanie2021recent}. In biosensing, nanocomposites can be engineered to create susceptible and selective sensors. Nanoparticles can act as signal amplifiers, enhancing the detection signal through their distinctive optical, electrical, or catalytic properties. The matrix material provides stability, mechanical strength, and a platform for biomolecular immobilization. By judiciously selecting nanoparticle types and incorporating them into the matrix, nanocomposite-based biosensors can achieve superior sensitivity, rapid response, and the capability to detect a wide range of analytes, including biomolecules and pathogens. For instance, incorporating hemin and silver-coated gold nanoparticles into a graphene oxide sheet led to a highly stable catalytic nanozyme with excellent detection performance \cite{kumar2017facile}. 

One-dimensional (1D) nanomaterials have nanoscale dimensions in two directions, while the third dimension is in the micrometer range. Carbon nanotubes (CNTs) and nanowires are noteworthy examples. CNTs are cylindrical nanostructures of carbon atoms arranged in a hexagonal lattice, forming a tubular shape. Due to their unique atomic arrangement, they exhibit remarkable mechanical, electrical, and thermal properties. CNTs can be single-walled (SWCNTs) or multi-walled (MWCNTs), with differing properties based on their structure \cite{tasis2006chemistry,maruyama2021carbon}. SWCNTs have extraordinary electrical conductivity and can be semiconducting or metallic, making them ideal for various electronic and energy storage applications. MWCNTs, on the other hand, possess exceptional strength and are used in reinforcement materials. Their high aspect ratio, surface area, and tunable properties have led to their utilization in diverse fields, including nanotechnology, materials science, electronics, and biomedical applications.

Two-dimensional (2D) nanomaterials have nanoscale dimensions in one direction while the other two remain relatively larger. The most notable example is graphene, a single layer of carbon atoms arranged in a two-dimensional honeycomb lattice. Graphene has garnered immense attention for its exceptional properties and diverse applications, particularly in biosensing \cite{pumera2011graphene,kuila2011recent,morales2017graphene}. Its remarkable electrical conductivity, high surface area, and biocompatibility make it a promising biosensor candidate. Graphene-based biosensors can detect biomolecules with high sensitivity and specificity, as the binding of target molecules leads to changes in their electrical properties. Its two-dimensional nature enables efficient interaction with biomolecules, enhancing sensor performance. Additionally, graphene's ease of functionalization allows the attachment of specific biomolecular recognition elements, enhancing selectivity \cite{kumar2016development,ghosal2018biomedical}. 

Three-dimensional (3D) nanomaterials are advanced structures that extend into the nanoscale in three spatial dimensions, offering unique properties and a high degree of control over their physical and chemical characteristics. These materials are recognized for their exceptional electroactive surface area, which allows for a higher loading capacity of recognition elements, such as antibodies or aptamers, thereby making them highly effective in targeting specific analytes, amplifying signals, and facilitating efficient biosensing with increased sensitivity and specificity. This category includes intricate hierarchical nanoscale structures and nanocomposites, which play a significant role in 3D materials \cite{byakodi2022emerging}. A notable example of 3D nanomaterials used in biosensing is the utilization of 3D graphene nanostructures. For instance, Chen et al.\cite{chen2018three} developed a three-dimensional electrochemical DNA biosensor utilizing silver nanoparticles decorated on a 3D graphene foam to detect CYFRA21-1 in lung cancer samples. Another study employed a graphene-metallic hybrid trimetallic nanoflower composite (3D GR/AuPtPd) to detect epidermal growth factor receptor (EGFR) ctDNA in human serum \cite{chen2021crispr}. Moreover, 3D hollow photoactive nanomaterials (such as Hollow CdS@Au nanospheres) have been instrumental in constructing multimodal biosensors for carcinoembryonic antigen detection, offering increased sensitivity through enhanced light capture attributed to their unique hollow nanostructures \cite{zhou2022biometric}. 

Other types of classifications of nanomaterials (e.g., organic, carbon, and inorganic) have been extensively discussed in several published articles \cite{ealia2017review,joudeh2022nanoparticle}. In biomedicine, nanomaterials have shown significant promise in drug delivery systems, where nanoparticles can be functionalized to carry therapeutic agents and selectively target specific cells or tissues. Additionally, nanomaterials are utilized in diagnostic imaging and biosensing applications, where their unique properties enable susceptible and specific detection of biological analytes. However, despite their promising advantages, nanomaterials also raise concerns regarding their potential toxicity and environmental impact \cite{elsaesser2012toxicology,sengul2020toxicity,yang2021nanoparticle}. Due to their small size, nanomaterials can easily penetrate biological barriers and interact with living organisms in ways that larger particles cannot. Therefore, extensive research is ongoing to understand and mitigate the potential risks of using nanomaterials.

Nanomaterials present a wealth of opportunities for groundbreaking innovations in diverse fields. Their unique size-dependent properties and versatility allow for tailoring material behavior to specific applications, leading to advances in electronics, medicine, energy, environmental remediation, and beyond. As nanotechnology continues to evolve, responsible and sustainable development of nanomaterials remains critical to ensure their safe and beneficial integration into various technological and biomedical applications.

\section{Organic Bioelectronic Devices}

\subsection{Organic Field-Effect Transistors (OFETs)}
Organic Field-Effect Transistors (OFETs) are semiconductor devices that utilize organic materials as the active channel to control the flow of charge carriers (electrons or holes) between the source and drain electrodes, modulated by an externally applied electric field at the gate electrode. OFETs have gained considerable attention recently due to their potential for low-cost, flexible, and large-area electronic applications, such as displays, sensors, and integrated circuits \cite{berggren2007organic,mei2013integrated}.

The basic structure of an OFET consists of three main components: the source, drain, and gate electrodes, all deposited on a substrate. The active channel material, typically an organic semiconductor, forms a thin film between the source and drain electrodes. A gate insulator layer separates The gate electrode from the channel material, often made of organic or inorganic dielectric \cite{horowitz1998organic,picca2020ultimately}. Figure~\ref{fig5}(a) illustrates the basic components of the OFET. The operation of an OFET relies on applying a gate voltage, which creates an electric field across the gate insulator and the channel material. This electric field either enhances or depletes the concentration of charge carriers in the channel, depending on the type of OFET (n-type or p-type). In an n-type OFET, the applied gate voltage increases the concentration of electrons in the channel, while in a p-type OFET, it increases the concentration of holes. The modulation of charge carriers in the channel material leads to a change in the conductivity between the source and drain electrodes. This change in conductivity is responsible for amplifying the input signal at the gate and producing a corresponding output signal at the drain, making OFETs function as amplifiers or switches.

One of the significant advantages of OFETs is their compatibility with low-cost, large-area manufacturing processes, such as solution-based deposition techniques like spin-coating or inkjet printing. The solution processability of organic semiconductors allows for the fabrication of flexible and stretchable devices on various substrates, including plastic and paper. The versatility of organic materials enables tailoring the active channel's electronic properties to specific application requirements. By modifying the molecular structure or introducing chemical dopants, researchers can optimize the charge transport behavior, charge carrier mobility, and overall device performance of OFETs.

OFETs find applications in various electronic devices, including electronic paper, flexible displays, RFID tags, biosensors, and logic circuits \cite{zhang2023flexible}. Additionally, OFET-based sensors have been developed for detecting various environmental and biological analytes, making them attractive for applications in healthcare, environmental monitoring, and point-of-care diagnostics. However, despite their advantages, challenges in OFET technology remain, such as improving charge carrier mobility, stability, and reproducibility \cite{chan2019motivation,wang2020challenges}. Researchers continue to explore novel materials, device architectures, and fabrication techniques to enhance the performance and reliability of OFETs, paving the way for their integration into next-generation electronics and wearable technologies.

\subsection{Organic Electrochemical Transistors (OECTs)}
OECTs are electronic devices that utilize organic materials to enable ion-mediated modulation of electrical conductivity. These transistors have gained significant attention due to their unique properties, such as low operating voltage, biocompatibility, and mechanical flexibility, making them suitable for various applications, including biosensing, neuromorphic computing, and bioelectronics \cite{ajayan2023organic}. The basic structure of an OECT consists of three main components: the source, drain, and gate electrodes, all integrated into a substrate \cite{friedlein2018device}. Figure~\ref{fig5}(b) shows a typical OECT schematic diagram. The operation of an OECT relies on the electrochemical doping and de-doping of the organic channel material. When a voltage is applied to the gate electrode, ions from the electrolyte solution penetrate the organic channel material, creating mobile charge carriers, either positively charged holes or negatively charged ions. This process is known as redox doping or ion-electron coupling. The presence of mobile charge carriers in the channel material modulates its electrical conductivity, affecting the current flow between the source and drain electrodes \cite{ait2021simple}. The channel's doping level can be adjusted by controlling the gate voltage, amplifying the input signal, and resulting in large changes in the output current \cite{bernards2007steady}. This unique ion-modulated transistor behavior sets OECTs apart from traditional field-effect transistors (FETs), where the current flow is regulated by applying an electric field across the gate-insulator interface.

The OECT devices work in two modes: depletion and accumulation modes \cite{rivnay2018organic}. By default, the depletion mode OECT operates with its channel in a conducting (ON) state, requiring an applied gate voltage to reduce its conductivity or switch it OFF. This type of organic transistor is constructed using organic semiconductor materials (e.g., PEDOT:PSS) and relies on ion transport within the semiconductor to modulate its conductivity. In contrast, the accumulation mode OECT remains in a non-conducting (OFF) state until a negative gate voltage is applied. This voltage accumulates charge carriers within the organic semiconductor channel, allowing current to flow and turning the transistor ON. Like Depletion Mode OECTs, Accumulation Mode OECTs also employ organic semiconductors (e.g., p(g2T-TT)) and ion transport for their operation, but their default state is non-conductive, requiring a gate voltage to activate them. 

One of the key advantages of OECTs is their biocompatibility, which enables their integration into biological systems without inducing significant adverse effects. This property makes OECTs ideal for interfacing with living cells and tissues, enabling applications in neural interfaces and bioelectronic devices \cite{chen2020recent}. Additionally, OECTs operate at low voltages, reducing power consumption and enabling the development of energy-efficient electronic systems. OECTs are widely employed in biosensing applications because they can transduce ion concentrations into electrical signals. By functionalizing the OECT channel with specific biomolecules or enzymes, the device can selectively detect and quantify target analytes, such as ions, neurotransmitters, glucose, or DNA, with high sensitivity and specificity. These biosensors find applications in medical diagnostics, environmental monitoring, and wearable health monitoring. Furthermore, OECTs have been utilized in neuromorphic computing, where they mimic the behavior of biological neurons in artificial neural networks. Their ion-mediated operation allows for dynamic signal processing and synaptic-like behavior, making them promising candidates for brain-inspired computing and pattern recognition tasks.

While OECTs offer numerous advantages, challenges remain in optimizing their stability, reproducibility, and scalability for large-scale production. Researchers continue to explore novel materials, device architectures, and fabrication methods to enhance the performance and reliability of OECTs, paving the way for their widespread adoption in cutting-edge electronic and bioelectronic technologies.

\subsection{Organic Electronic Ion Pumps (OEIPs)}
\begin{figure}
\centering
\includegraphics[width=\textwidth]{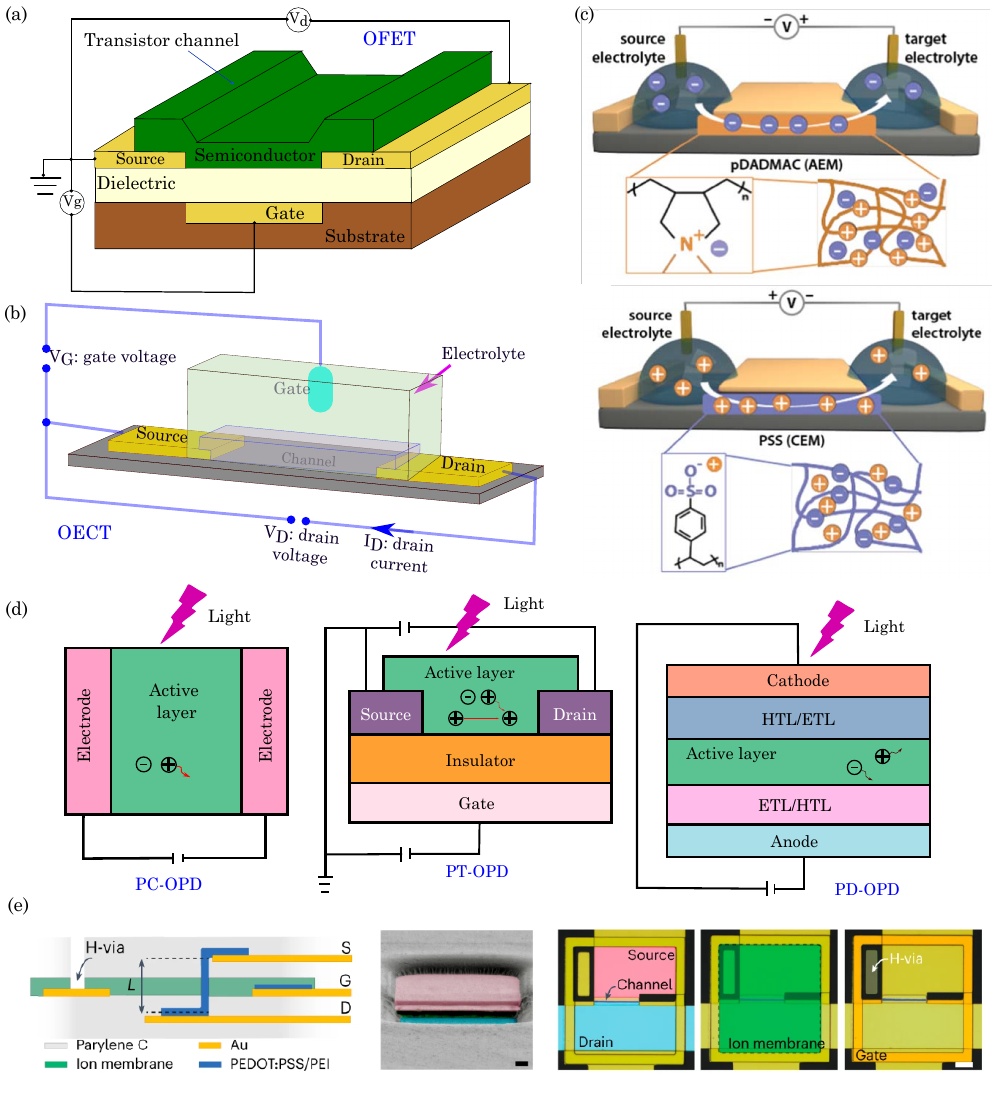}
\caption{\textbf{(a)} Schematic diagram of organic field-effect transistors (OFETs); \textbf{(b)} Typical structure of an organic electrochemical transistor (OECT). Adapted from Friedlein et al. \cite{friedlein2018device}, \emph{Organic Electronics 2018, 63, 398–414}, \copyright2018 The Authors, licensed under a Creative Commons license.; \textbf{(c)} Schematic illustration of OEIP device configuration and the working principle of a potential ion-selective OEIP (top) and a cation-selective OEIP (bottom). As illustrated, applying a potential between electrodes establishes an electrochemical circuit. Within this circuit, cations or anions from a source electrolyte are selectively conveyed to a target electrolyte through an ion exchange membrane. Adapted with permission from Cherian et al. \cite{cherian2019large}, \emph{Flex. Print. Electron. 2019, 4, 02200}. \copyright2019 The Authors, published by IOP Publishing Ltd. under the terms of the Creative Commons Attribution 3.0 license.; \textbf{(d)} Device configurations of OPDs: organic photoconductor (PC-OPD), organic photoresistor (PT-OPD), and organic photodiode (PD-OPD). \emph{ETL}- electron transport layer, and \emph{HTL}- hole transport layer. Adapted with permission from Liu et al.\cite{liu2020organic}, \emph{Solar Rrl 2020, 4, 7, 2000139}, \copyright2020 WILEY‐VCH Verlag GmbH \& Co. KGaA, Weinheim.; \textbf{(e)} Schematic of a vIGT (left), L: vertical channel length, S: source, G: gate, D: drain, Colourized cross-section scanning electron microscopy image of a vIGT (center). The pink and blue regions are the source and drain contacts, respectively, and the optical micrograph displays the top view of an individual vIGT (right), blue color: drain contact, pink:  source contact, green: ion membrane.  Reprinted with permission from Cea et al. \cite{cea2023integrated}, \emph{Nat. Mater. 2023, 22, 1227–1235}. \copyright2022 The Authors, published by Springer Nature under a Creative Commons Attribution 4.0 International License.}
\label{fig5}
\end{figure}

Organic electronic ion pumps represent a burgeoning area of research in bioelectronics, where the principles of organic materials and electronics converge to create advanced systems for ion transport \cite{simon2016organic,mei2022bioinspired}. These ion pumps utilize organic materials with specific ion-selective properties to enable controlled and precise transport of ions, such as cations, anions, protons, or other charged species. The underlying principle involves utilizing organic materials that can change their state, conductivity, or permeability when exposed to external stimuli such as voltage or chemical signals. By applying an electrical potential, these materials can effectively regulate the movement of ions across a membrane or interface. Figure~\ref{fig5}(c) depicts the typical device configuration of a potential ion-selective OEIP and a cation-selective OEIP.  As shown, an OEIP comprises two electrodes separated by an ion-exchange membrane.  When a voltage is applied between the two electrodes, one of which is positioned beneath the ion reservoir and the other situated at the target area, cations (or anions) migrate from the reservoir through the respective exchange membrane to the delivery site \cite{ohayon2020organic}.

The OEIPs' capability to manipulate ion transport holds significant implications across diverse domains, ranging from addressing therapeutic challenges through targeted drug delivery and neural modulation to applications in biotechnology and bioengineering. Examples of OEIPs encompass triggering cell signaling in vitro \cite{isaksson2007electronic,tybrandt2009translating}, controlling epileptiform activity in brain slice models \cite{proctor2018electrophoretic}, influencing sensory functions in vivo \cite{simon2009organic}, serving as pain therapy in awake animals \cite{jonsson2015therapy}, and even regulating plant growth through the delivery of phytohormones \cite{poxson2017regulating}.

Organic electronic ion pumps offer several advantages, including biocompatibility, flexibility, and the potential for miniaturization. These properties make them well-suited for integration into bioelectronic devices and implantable systems \cite{jakevsova2019wireless,strakosas2021electronic}. OEIPs can be designed to work in tandem with other components like sensors, actuators, and communication modules. This integration allows for dynamic feedback loops, enabling real-time adjustments in ion transport based on physiological responses or external triggers. As research advances, the development of organic electronic ion pumps has the potential to revolutionize the field of bioelectronics, opening up new avenues for creating smart and responsive bio-integrated systems that interface seamlessly with biological environments and hold promise for a range of medical, therapeutic, and biotechnological applications.

\subsection{Organic Photodetectors (OPDs)}
Organic photodetectors (OPDs) are optoelectronic devices that convert incident light into electrical signals through the photoelectric effect, utilizing organic materials as the active absorbing layer. These devices have gained significant attention due to their potential for low-cost, flexible, and large-area optoelectronic applications, including image sensors, photodiodes, and light detectors \cite{yang2019development}.

The basic structure of an organic photodetector typically comprises an organic semiconductor layer sandwiched between two electrodes, acting as the anode and cathode. The organic semiconductor layer absorbs photons from incident light, generating electron-hole pairs in the material. Depending on the type of OPD, either electrons or holes are transported through the organic layer to the respective electrodes. The operation of an OPD is based on the efficient generation, separation, and collection of photo-excited charge carriers. When photons with energy equal to or greater than the semiconductor bandgap are absorbed, excitons (electron-hole pairs) are created. These excitons must be efficiently dissociated into free-charge carriers to generate a photocurrent. To enhance exciton dissociation, OPDs often incorporate donor-acceptor heterojunctions, where the energy levels of the donor and acceptor materials promote efficient charge separation. The photocurrent generated in the organic layer is collected at the electrodes, and the magnitude of the photocurrent is proportional to the intensity of the incident light. Figure~\ref{fig5}(d) shows the three different architectures of OPDs, namely, photoconductors‐based OPDs (PC‐OPDs), phototransistors‐based OPDs (PT‐OPDs), and photodiodes‐based OPDs (PD‐OPDs). The PT‐OPDs comprised three electrodes: gate, source, and drain. In contrast, PC-OPDs and PD-OPDs are configured based on two electrodes (i.e., anode and cathode). PC-OPDs leverage photoconductivity in organic materials to detect light, offering sensitivity across a wide spectrum. PT-OPDs employ a transistor structure for amplified sensitivity, making them ideal for low-light conditions. PD-OPDs combine organic semiconductors with photodiode principles, delivering high-speed and efficient light detection, which is crucial for applications like optical communication and rapid imaging. Each OPD type caters to specific needs, providing a versatile toolkit for various optoelectronic applications.

Apart from Sandwich types, planar-type photodetectors have also been used. These photodetectors are semiconductor devices with a planar structure designed for efficient light detection and conversion into electrical signals. These devices, typically made from semiconductor materials like silicon (Si) and gallium arsenide (GaAs) \cite{schmidt1983position,tull2002new,caria2002gallium}, operate on a fundamental principle where incident photons with energy greater than the semiconductor's bandgap generate electron-hole pairs when they strike the device's surface. The resulting free electrons and holes are then separated and collected by an internal electric field, creating a photocurrent or a change in voltage, which is directly proportional to the intensity of the incident light. Planar-type photodetectors encompass various designs, including photodiodes, phototransistors, and avalanche photodiodes \cite{li2023tricolor,liu2018fabrication,martyniuk2023infrared}. Photodiodes collect the separated carriers directly, offering a linear response to incident light. Phototransistors amplify the signal by using the generated carriers to control a larger current flow, while avalanche photodiodes, intended for applications requiring high sensitivity, leverage avalanche multiplication to produce a substantial number of charge carriers. These photodetectors are extensively applied in optical communication systems, imaging devices, optical sensors, and many applications demanding light detection.

OPDs exhibit high responsivity, low dark current, and fast response times, making them suitable for a wide range of light detection applications. One of the key advantages of OPDs is their compatibility with solution-based deposition techniques, enabling the fabrication of large-area and flexible devices on various substrates. The tunability of organic materials allows for optimizing their light absorption properties to match specific wavelengths or spectral ranges, making OPDs versatile for various optical sensing and imaging applications. OPDs are used in diverse optoelectronic devices, such as image sensors \cite{ng2008flexible,eckstein2018fully,song2022doping}, light-sensitive arrays, photodetector arrays \cite{calvi2022highly}, and position-sensitive detectors \cite{rand2003thin,cabanillas2011organic,li2023bioadhesive}. They find applications in digital cameras, medical imaging, light-based communication systems, and optical sensors for environmental monitoring and industrial applications. Additionally, organic materials' flexibility and lightweight nature enable the development of wearable and conformable photodetectors for wearable health monitoring, biometric sensing, and smart textiles.

Despite their advantages, challenges in OPD technology include improving the external quantum efficiency, enhancing the stability of organic materials under prolonged light exposure, and achieving high-speed response times for rapid optical sensing applications \cite{liu2020organic}. Researchers are actively exploring novel organic materials, device architectures, and engineering strategies to overcome these challenges and unlock the full potential of organic photodetectors in the emerging field of organic optoelectronics.

\subsection{Organic Bioelectronic Implants}
Organic bioelectronic implants are advanced medical devices integrating organic electronic materials and components into living tissues to enable various therapeutic or diagnostic functionalities. These implants represent a cutting-edge field of research and development in the intersection of organic electronics and biomedicine, offering unique advantages for medical applications \cite{mariello2022recent,wu2023tissue}.

Organic bioelectronic implants constitute a complex assembly of crucial components aimed at interfacing with biological systems while delivering therapeutic or monitoring functions. Central to their design are organic semiconductors, conductive polymers like poly(3,4-ethylene dioxythiophene): poly(styrene sulfonate) (PEDOT:PSS), and specialized organic electronic materials meticulously chosen for their biocompatibility, mechanical flexibility, and ability to seamlessly integrate with biological tissues, all while evading significant immune responses \cite{jonsson2015therapy}. These materials serve as the foundation for the implant's active elements.

Organic bioelectronic implants exhibit adaptability by incorporating sensors to monitor vital physiological parameters like pH, temperature, glucose levels, or specific biomarkers. Additionally, they integrate stimulating components such as electrodes or transducers capable of delivering targeted electrical or chemical signals. These signals serve therapeutic objectives such as deep brain stimulation or promoting neural regeneration. To ensure the longevity and efficacy of the implant, it is encapsulated within biocompatible materials or coatings. This encapsulation acts as a protective barrier against unwanted interactions with the surrounding biological environment. In a recent example, Cea et al. \cite{cea2023integrated} developed a tiny, fully organic bioelectronic device that acquires and transmits brain signals and self-powers. The device is about 100 times smaller than a human hair and is based on an IGT (internal-ion-gated organic electrochemical transistor) architecture, the vIGT (vertical internal ion-gated organic electrochemical transistor) that incorporates a vertical channel made of PEDOT:PSS and a miniaturized water conduit (H-via) from the surface of the device through the ion membrane layer to permit channel hydration, demonstrating long-term stability, high electrical performance, and low-voltage operation to prevent biological tissue damage. Figure~\ref{fig5}(e) demonstrates the device architecture schematics, SEM image and optical micrograph of the vIGT.  

Furthermore, these implants harness wireless communication, enabling connectivity with external devices for data collection, remote control, and programming. This breakthrough promises a revolution in patient monitoring and treatment optimization, as demonstrated by recent studies \cite{ferguson2011wireless, khan2022recent, tian2023implant}. They also employ innovative power management systems, including energy harvesting and wireless charging, ensuring sustainable operation and reducing the need for frequent battery replacements. An additional advantage lies in the mechanical flexibility of organic bioelectronic implants, enabling seamless integration with irregular and dynamic tissue shapes and movements. This adaptability proves invaluable when implants are placed in soft, curved body regions like the brain, heart, or spinal cord.

Moreover, recent advancements have led to the development of biodegradable organic bioelectronic implants. These designs gradually dissolve over time, minimizing harm to surrounding tissues and eliminating the need for additional surgical removal. These implantable bioelectronic devices offer immense potential across diverse medical applications. Organic sensors can precisely monitor drug release rates and tailor dosages for personalized drug administration. Moreover, they facilitate tissue regeneration by offering electrical or biochemical cues to spur cell growth and tissue repair. Notably, these devices find application in neuroprosthetics, including cochlear implants for hearing restoration and retinal implants for vision enhancement \cite{dimov2022semiconducting}. Additionally, they are employed for simulating peripheral nerves to treat disorders resistant to traditional pharmacological interventions. 

As organic bioelectronic implants advance, ongoing research focuses on optimizing biocompatibility, stability, and long-term functionality and addressing challenges related to immune responses, long-term biointegration, and regulatory approvals. With continued innovations, organic bioelectronic implants have the potential to revolutionize personalized medicine, ushering in a new era of advanced healthcare and improved quality of life for patients.

\section{Fabrication methods}
Organic bioelectronic devices are predominantly fabricated/patterned using several approaches, such as organic thin-film deposition methods, patterning techniques, 3D printing, and organic synthesis. 

\begin{figure}
\centering
\includegraphics[width=\textwidth]{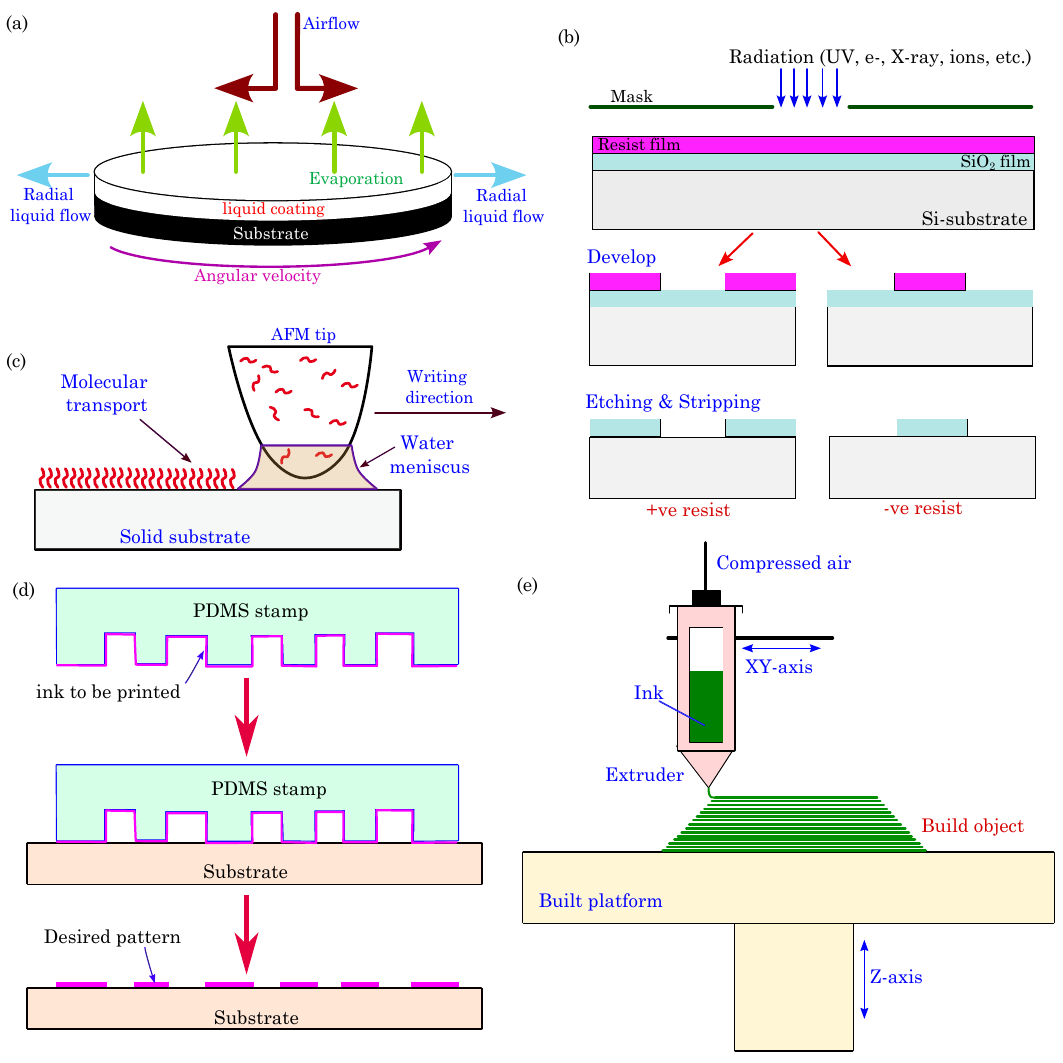}
\caption{Schematic diagram of various fabrication methods: \textbf{(a)} spin-coating process; \textbf{(b)} photolithography ; \textbf{(c)} dip-pen nanolithography (DPN); \textbf{(d)} Micro contact printing ($\mu$CP); \textbf{(e)} Direct ink writing (DIW).}
\label{fig6}
\end{figure}

\textbf{Organic thin-film deposition}: These methods are widely used for depositing thin films of organic materials on substrates with controlled thickness and uniformity. One common technique is spin-coating. In spin-coating, an organic material solution, such as semiconductors, conductive polymers, or other active components, is deposited onto a flat substrate, typically a silicon wafer or glass, which can be further integrated into a device (see schematics in Figure~\ref{fig6}a). As the substrate spins at high speeds, centrifugal forces evenly distribute the material, resulting in a thin, uniform film. The spin coating offers precise control over the thickness and quality of the deposited organic films, allowing researchers to optimize these devices' electrical and optical properties. This method is ideal for producing organic semiconductor layers used in devices like organic field-effect transistors (OFETs) and organic photodetectors \cite{liu2012forming,xie2019highly}. With its scalability, cost-effectiveness, and ongoing refinements, spin coating plays a central role in various applications, from flexible electronics to medical diagnostics and wearable health monitoring, ensuring the advancement of organic bioelectronics in diverse fields. 

Vacuum evaporation is another thin-film deposition method. It facilitates the precise deposition of organic materials onto various substrates under reduced pressure conditions. In this process, organic materials, such as semiconductors, conductive polymers, and other key bioelectronic components, are heated to their vaporization points and then allowed to condense onto the target substrate, creating thin organic films with exceptional uniformity and precise thickness control. This level of control is indispensable in developing organic electronic devices, including organic field-effect transistors (OFETs) and organic photodetectors, where the properties of the organic layer directly influence device performance. Vacuum evaporation enables the sequential deposition of multiple organic layers, making it possible to design complex device architectures. This capability is invaluable as organic bioelectronic devices often require distinct functional layers for sensing, signal processing, and data transmission. Additionally, vacuum evaporation is a low-temperature deposition technique that safeguards the structural integrity of heat-sensitive organic materials. It also provides a pristine vacuum environment that minimizes contamination, ensuring the quality of the deposited organic films. In the realm of organic bioelectronics, vacuum evaporation plays a critical role in manufacturing devices like biosensors, organic photovoltaics, and implantable bioelectronic systems. For instance, vacuum evaporation is often used in OLED fabrication \cite{cho2020solution,jung2021patternable}. 

\textbf{Patterning techniques}: Several methods are employed for patterning electrodes for organic electronic devices, such as photolithography, e-beam lithography, dip-pen lithography, inkjet printing, micro-contact printing, screen printing, direct ink writing, laser writing, etc.

\emph{Photolithography}: It is a well-established technique for patterning organic materials at micron and submicron scales. The photolithography process begins with a substrate, typically made of silicon or glass, coated with a layer of photoresist, a photosensitive organic material. A photomask containing the desired pattern is placed near the photoresist-coated substrate,  and the entire assembly is exposed to ultraviolet (UV) light. The exposed regions undergo a chemical change, making them either more soluble (in the case of positive photoresists) or less soluble (for negative photoresists) in a developer solution, depending on the type of photoresist used. The developer solution is applied to the substrate, removing the selected areas and leaving behind the desired pattern (see schematics in Figure~\ref{fig6}b). Photolithography stands out due to its exceptional resolution and accuracy, making it capable of crafting intricate micro and nanoscale structures relevant to organic bioelectronics. The adaptability of this technique to a range of organic materials facilitates the fabrication of diverse bioelectronic components. Nonetheless, cautious handling of organic materials is essential, as some may be sensitive to UV exposure and chemical developers. Additionally, meticulous design of photomasks is imperative to achieve the desired patterns. 
Photolithography is employed in organic bioelectronic device fabrication to create features like electrodes, sensor structures, and microfluidic channels \cite{defranco2006photolithographic,khodagholy2011highly,sessolo2013easy,nawaz2021organic}.

\emph{Electron beam (e-beam) Lithography}: e-beam lithography or EBL is an advanced nanofabrication technique that operates on the fundamental principle of using a focused beam of electrons to create incredibly fine patterns and structures at the nanometer scale.  It has found applications in various fields, including semiconductor device fabrication, nanotechnology, and micro-electromechanical systems (MEMS). Unlike conventional photolithography, EBL can achieve unparalleled resolution, crafting features with dimensions down to just a few nanometers. This capability arises from its direct-write process, where a precisely controlled electron beam moves across an electron-sensitive resist on a substrate to define intricate custom patterns. While slower and more complex than some alternatives, e-beam lithography is crucial in developing advanced nanoscale devices, specialized structures in research laboratories, and creating masks and photomasks for semiconductor manufacturing. 

\emph{Dip-Pen Nanolithography (DPN)}: DPN is an advanced nanofabrication technique that leverages the precision of scanning probe microscopy (SPM) for the controlled deposition of molecules, nanoparticles, or biomolecules onto a substrate with nanometer-scale precision. In this method, an atomic force microscope (AFM) tip coated with an "ink" material is submerged, or "dipped," into the ink and then brought into contact with a substrate under the guidance of the AFM (schematics in Figure~\ref{fig6}c). DPN is renowned for its extraordinary sub-10 nanometer resolution, making it a pivotal tool in various domains, such as nanoelectronics, nanophotonics, and nanobiotechnology. Its remarkable versatility extends to the patterning of diverse materials, including conducting polymers, biological compounds like proteins or DNA, nanoparticles, and more, enabling the creation of various structures, from lines and dots to intricate two-dimensional and three-dimensional designs. DPN finds applications across several domains: in nanoelectronics for the development of nanoscale electronic components and features on semiconductor chips, in nanophotonics for crafting optical devices, photonic circuits, and metamaterials, in biosensing for creating highly sensitive and specific biosensors, in surface functionalization for engineering specific surface properties, and in nanomaterials synthesis for precise control of nanoparticle properties. While DPN offers exceptional precision, it can be a relatively slow and serial process, limiting its application for large-scale manufacturing, and the choice of ink, tip, substrate, and environmental conditions significantly influence pattern quality and reproducibility.

\emph{Micro contact printing ($\mu$CP)}: $\mu$CP is a widely used soft lithography technique employed for precise and controlled deposition of materials, often in the form of self-assembled monolayers (SAMs), on a substrate. The process is akin to conventional rubber stamp printing but on a micro- and nanoscale. $\mu$CP employs an elastomeric stamp, usually made of polydimethylsiloxane (PDMS), engineered with relief microstructures or patterns on its surface. The stamp is coated with an "ink" or material, which adheres only to the relief patterns. The stamp is then gently brought into contact with a substrate, transferring the material onto the substrate in the desired pattern (see schematics in Figure~\ref{fig6}d). This process offers several advantages, including simplicity, cost-effectiveness, and the ability to create well-defined and precisely placed chemical patterns on various substrates, including metals, semiconductors, and organic materials. $\mu$CP is particularly valuable for creating surface chemistry modifications and developing biomolecule arrays and microscale patterning for various applications, including biosensors, microelectronics, and microfluidics. However, $\mu$CP also has some limitations. It may be less suitable for large-scale or high-throughput manufacturing processes, as it is inherently a serial process. The resolution of $\mu$CP depends on the stamp's relief structures and may not achieve the sub-10-nanometer scale of some advanced lithography techniques. Additionally, controlling the uniformity of the ink layer and ensuring consistent contact between the stamp and substrate can be challenging. Despite these limitations, micro-contact printing remains a powerful tool for many micro- and nanofabrication applications, particularly in research and prototyping scenarios.

\emph{Inkjet printing}: Inkjet printing, a highly versatile technique, has become integral to the realm of organic bioelectronics. The process involves depositing minuscule ink droplets onto a substrate, enabling controlled patterning of various functional materials, including organic semiconductors, conductive polymers, and biologically relevant molecules. Its prominence in this field stems from multiple advantages, such as exceptional precision and resolution, broad material compatibility, reduced material wastage due to its additive nature, high levels of customization to adapt complex designs for specific applications, non-contact printing, and scalability that accommodates everything from research-level prototyping to large-scale production \cite{setti2007hrp,weng2012inkjet,greco2013patterned,kumar2019additive}. Inkjet printing plays a pivotal role in fabricating components for organic bioelectronic devices, including sensors, transistors, and electrochemical systems, and excels in the precise deposition of biomolecules crucial for biosensing and detection applications. This technology is a cornerstone in developing advanced medical diagnostics, wearable health monitoring devices, and implantable bioelectronics, promising significant contributions to healthcare and environmental monitoring.

\emph{Laser Writing}: Laser writing, also known as laser-induced forward transfer (LIFT), is an advanced microfabrication technique that employs a high-intensity laser beam to transfer material from a donor layer to a receiver substrate, enabling the precise deposition of micro- or nanoscale features. A laser pulse generates a shockwave within the donor material, propelling a small amount of material toward a transparent receiver substrate placed above it. This method offers exceptional precision, allowing for fine control over the position and size of the deposited material, making it ideal for creating intricate patterns, microarrays, and electronic devices. One of its significant advantages is versatility, as it can be used with various materials, including organic polymers, conductive substances, and biological compounds, making it suitable for applications ranging from organic electronics to biosensors. Due to its non-contact nature and direct-write capabilities, laser writing is precious for handling sensitive materials and enabling rapid prototyping. With the potential to achieve sub-micron resolutions, this technique has widespread applications in microelectronics, flexible electronics, organic photovoltaics, microfluidics, and tissue engineering, where high-resolution and customized structures are paramount for research, development, and specialized manufacturing processes.

\textbf{3D printing}: The field of bioelectronics has witnessed remarkable progress with the integration of 3D printing technologies. These technologies are known for their streamlined processes, which empower the creation of intricate three-dimensional structures with exceptional precision, scalability, and adaptability \cite{zhang2019recent,sreenilayam2020advanced,saadi2022direct}. Various 3D printing techniques, including fuse deposition modeling (FDM), stereolithography (SLA), digital light processing (DLP), selective laser sintering (SLS), and direct ink writing (DIW), have been instrumental in patterning and fabricating materials with diverse strategies. Nevertheless, many of these technologies are often associated with specific material classes, such as thermoplastic polymers for FDM, photopolymer resins for SLA and DLP, and powdered polymers or metals for SLS, which impose limitations on the customization of inks. Within this landscape, DIW, an extrusion-based 3D printing technique that constructs 3D structures layer-by-layer through the precise deposition of inks via fine nozzles ((schematics in Figure~\ref{fig6}e)), has emerged as the most versatile 3D printing technology, offering unprecedented capabilities for the development of bioelectronics. These inks may encompass various materials, spanning metals, ceramics, polymers, carbons, and even biocompatible substances such as cells or gels. The DIW printer follows a computer-generated design to create intricate and customized objects layer by layer \cite{tay2023direct}.

\textbf{Chemical methods}: Organic bioelectronic devices can also be fabricated through diverse chemical methods, including polymerization, chemical vapor deposition (CVD), and self-assembly \cite{choi2002bioelectronic,iost2012layer,wang2017cvd,heydari2019device,torricelli2021electrolyte,wang2022flexible,balakrishnan2022recent}. These methods allow for precise control over the molecular structure of materials, enabling the design of custom organic semiconductors, conductive polymers, and biocompatible coatings. Polymerization involves the creation of organic materials through the reaction of monomers, resulting in polymers with desired properties. CVD entails depositing thin films of organic materials from vapor-phase precursors, ensuring uniform and controlled material growth. Self-assembly allows organic molecules to spontaneously arrange into ordered structures, which can be fine-tuned for targeted functionalities \cite{schwartz2001mechanisms,kwon2010nanoscale}. 

Table~\ref{tab1} summarizes the fabrication techniques used for fabricating organic bioelectronic devices. These fabrication methods provide versatility in designing organic bioelectronic materials with unique characteristics, such as high sensitivity, flexibility, and biocompatibility. By leveraging organic thin-film deposition and organic synthesis techniques, researchers can engineer materials tailored to the requirements of biosensing, medical diagnostics, and wearable health monitoring applications, among others. Continued advancements in organic bioelectronic material fabrication hold great potential in revolutionizing the landscape of bioelectronics and contributing to breakthroughs in medical technologies and personalized healthcare.

\begin{table} 
 \caption{Summary of fabrication techniques for organic electronics.} 
 \label{tab1}
 \begin{tabular} [c]{p{3.5cm}p{7cm}p{3.5cm}}
\toprule
\textbf{Fabrication techniques}	& \textbf{material}	& \textbf{References} \\
 \midrule
Spin coating & 2D crystalline film from 2,7-diocty[1]benzothieno[3,2-b]benzothiophene (C8-BTBT), PDMS, organic semiconductor films, PEDOT:PSS    & \cite{dai2019fabrication,gablech2023state,wang2020directly,chakraborty2023conductive} \\
Photolithography & PEDOT:PSS, OLED   &\cite{defranco2006photolithographic,ouyang2015surface,dadras2022multiphoton,son2020descumming} \\
E-beam lithography & PPy, poly(chloro-p-xylylene) (Parylene C), Biomolecules   & \cite{donthu2005facile,gomez2007micropatterned,scholten2016electron,kolodziej2012electron} \\
Dip-pen nanolithography & sulfonated polyaniline (SPAN), PPy, PEDOT, ferroelectric copolymer poly (vinylidene fluoride–
trifluorethylene)   & \cite{lim2002electrostatically,tang2004fabrication}\\
Inkjet printing & PEDOT:PSS, PPy  & \cite{setti2007hrp,weng2012inkjet,greco2013patterned,eom2009polymer} \\
Micro contact printing & PPy, PEDOT, Proteins, Ultrathin Gate
Dielectrics,  alkyl and fluoroalkylphosphonic acid  &  \cite{park2014hydrogel,ricoult2014humidified,hager2022microcontact,zschieschang2008microcontact,hirata2015high}\\
Laser writing & PEDOT, PANI, laser-induced porous graphene  & \cite{liu2020laser, zhu2021portable} \\
Direct ink writing & PEDOT:PSS, PEDOT:PSS-PEO, holey graphene oxide (hGO), eutectic gallium–indium (EGaIn)-based liquid metal embedded elastomers, AgNPs, MWCNT, rGO/CNT, silicone   & \cite{yuk20203d,plog2023extremely,lacey2018extrusion,kee2020direct,won20223d,wang2022phase,tang2018generalized} \\
Chemical vapor deposition & Poly(p-xylylene), PEDOT  & \cite{khlyustova2020vapor,wang2018high} \\
 \bottomrule
\end{tabular}
\end{table}

\section{Biosensing Mechanisms}
\begin{figure}
\centering
\includegraphics[width=\textwidth]{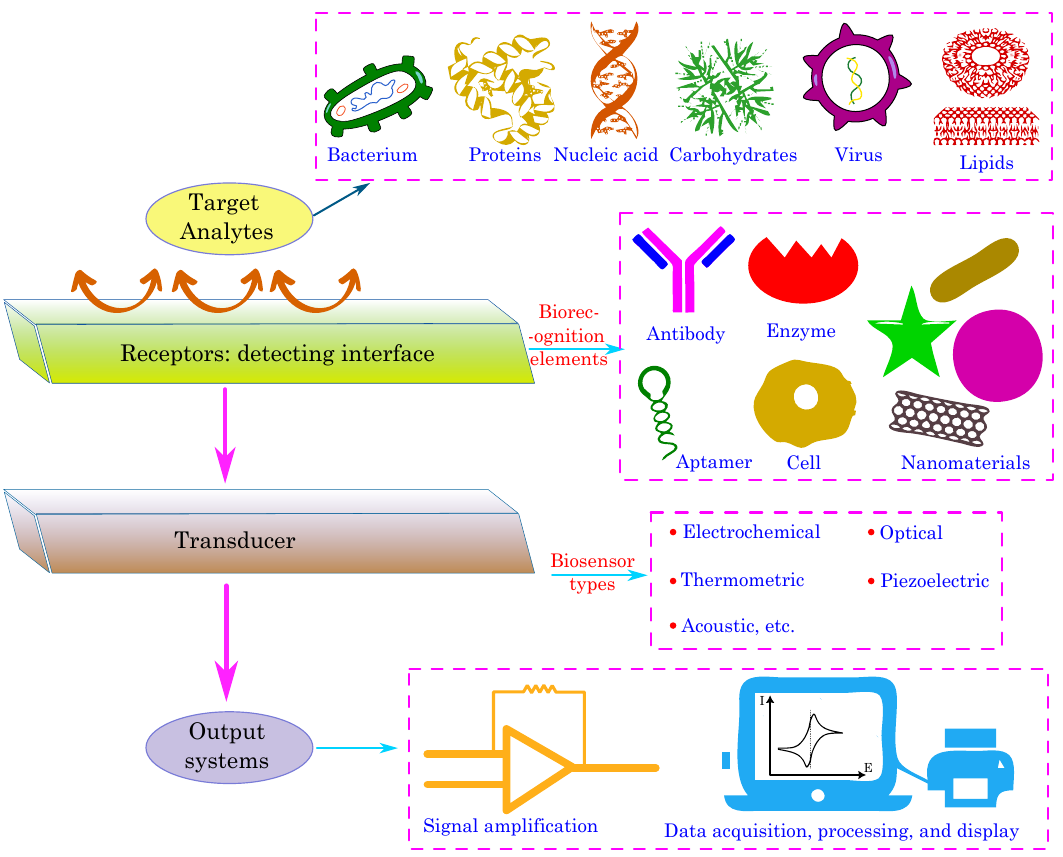}
\caption{Schematic illustration of key components of a typical biosensor.}
\label{fig7}
\end{figure}

A typical biosensor comprises several fundamental components: the target analytes, receptors or biorecognition elements, a transducer, and output systems \cite{macchia2020amplification,naresh2021review}. The target analyte is the specific substance under investigation, such as glucose, ammonia, alcohol, or lactose. Bioreceptors are biomolecules or biological entities capable of recognizing and binding to the target analyte. Examples of biorecognition components include enzymes, cells, aptamers, DNA/RNA strands, and antibodies. The role of the transducer is to convert the biorecognition event into a measurable signal, typically in the form of an electrical signal, which correlates with the quantity or presence of the chemical or biological target. This conversion process is known as signalization. Transducers generate optical or electrical signals directly corresponding to the interactions between analytes and bioreceptors. Finally, output systems encompass signal processing, amplification, and display units, facilitating the interpretation and presentation of the biosensor's results. Figure~\ref{fig7} illustrates the components of the typical biosensor.

\subsection{Electrochemical Sensing}
\begin{figure}
\centering
\includegraphics[width=\textwidth]{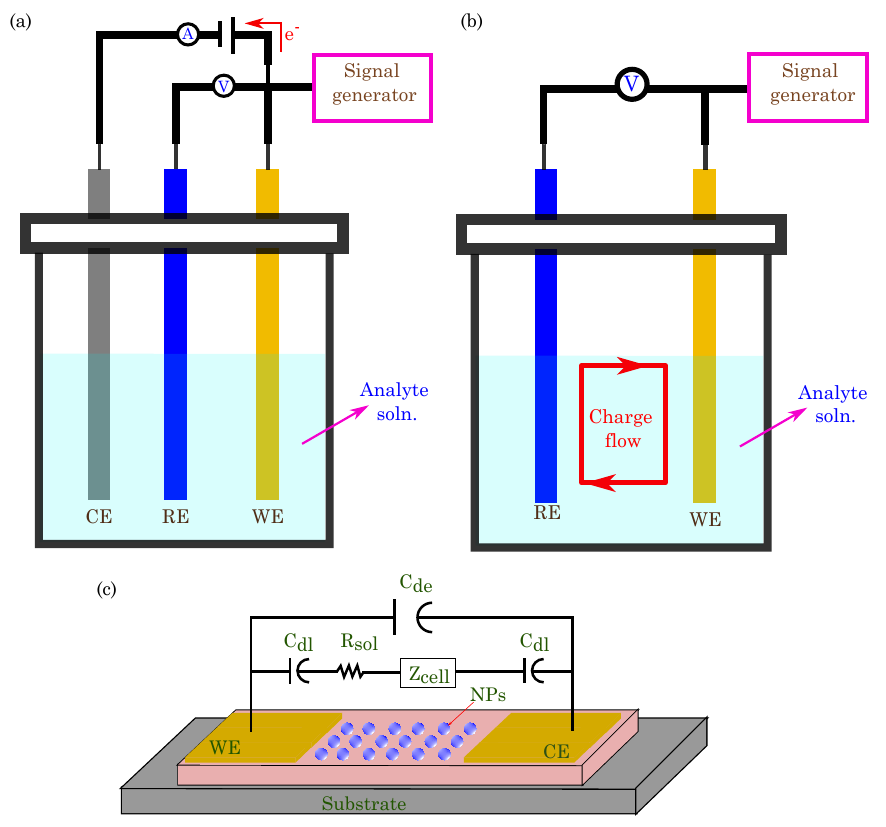}
\caption{Schematics configuration of different types of electrochemical sensors. \textbf{(a)} amperometric/voltammetric biosensor, \textbf{(b)} potentiometric biosensor, \textbf{(c)} impedimetric biosensor (Cdl = double-layer capacitance of the electrodes, Rsol = resistance of the solution, Cde = capacitance of the electrode, Zcell = impedance introduced by the bound nanoparticles). Adapted from Naresh and Lee \cite{naresh2021review}, \emph{Sensors, 2021, 21, 4, 1109}, \copyright2021 MDPI.}
\label{fig8}
\end{figure}

Electrochemical sensing is a powerful mechanism utilized in organic bioelectronics for detecting and quantifying various biomolecules and chemical species. This sensing platform measures electrical signals generated during electrochemical reactions at the interface between the organic material and the target analyte. Organic electrochemical sensors offer high sensitivity, rapid response times, and excellent selectivity, making them valuable medical diagnostics, environmental monitoring, and point-of-care testing tools. The fundamental principle behind electrochemical sensing in organic bioelectronics lies in the redox properties of organic materials, which can undergo reversible electron transfer reactions \cite{grieshaber2008electrochemical,ronkainen2010electrochemical}. These redox-active organic materials, such as conducting polymers, redox enzymes, or organic nanoparticles, are integrated into the sensing platform to act as the transducer element. Electrochemical sensing involves two main components: electrode and redox reaction with the target analyte. The sensing platform typically comprises working (or indicator electrode), reference, and counter electrodes (in some cases, the two-electrode system can be used for electrochemical sensing) \cite{cui2015electrochemical,shanbhag2023fundamentals}. The working electrode (WE) is coated with the redox-active organic material, where the electrochemical reaction with the target analyte occurs. The reference electrode (RE) maintains a constant potential against which the working electrode's potential is measured. The counter electrode (CE) completes the electrical circuit and balances the current flow during the electrochemical reaction. When the target analyte comes into contact with the redox-active organic material on the working electrode, it induces an electrochemical reaction. The redox-active organic material is reversibly oxidized or reduced, transferring electrons between the analyte and the electrode surface. This electron transfer generates an electrical signal, such as a current or potential, which is proportional to the concentration of the target analyte. Different electrochemical sensing modalities employed in organic bioelectronics include:

\textbf{Amperometric Sensing}: Amperometric biosensors are a type of electrochemical biosensor used for quantitatively detecting and analyzing biological analytes. These biosensors rely on the measurement of current generated from an electrochemical redox reaction at the sensor's working electrode surface when the target analyte interacts with a biorecognition element (such as enzymes, antibodies, or nucleic acids) immobilized on the electrode. The basic setup of an amperometric biosensor typically consists of three main components: the working electrode, the reference electrode, and the counter electrode. The biorecognition element is immobilized in the working electrode, and the redox reaction occurs upon the target analyte's binding. The reference electrode maintains a constant potential, while the counter electrode completes the electrical circuit, allowing the flow of electrons during the redox reaction. When the target analyte binds to the biorecognition element on the working electrode's surface, it triggers the redox reaction, producing or consuming electroactive species (e.g., hydrogen peroxide or oxygen). The current generated from this redox reaction is directly proportional to the concentration of the target analyte in the sample. As the concentration of the analyte changes, the current also varies, providing quantitative information about the analyte concentration. Figure~\ref{fig8}(a) shows the schematics of amperometric-based biosensors.

\textbf{Voltammetric biosensing}: Voltammetric biosensors are a type of electrochemical biosensor that relies on the measurement of current as a function of an applied voltage or potential at the sensor's working electrode. These biosensors use the principles of voltammetry to detect and quantify the target analyte in a sample. The basic setup of a voltammetric biosensor includes a working electrode coated with a biorecognition element, a reference electrode, and a counter electrode. When an increasing or decreasing voltage is applied to the working electrode, a redox reaction occurs at the electrode surface, involving the oxidation and reduction of electroactive species. In the presence of the target analyte, the biorecognition element at the working electrode surface interacts with the analyte, leading to changes in the redox reaction of the electroactive species. These changes result in variations in the current measured at the working electrode, which can be correlated with the concentration of the target analyte in the sample.

\textbf{Potentiometric Sensing}: Potentiometric biosensors are a type of electrochemical biosensor used for the quantitative detection and analysis of biological analytes. Unlike amperometric biosensors that measure the current generated from a redox reaction, potentiometric biosensors rely on measuring potential or voltage changes at the sensor's working electrode surface when the target analyte interacts with a biorecognition element. The basic setup of a potentiometric biosensor includes a working electrode and a reference electrode (Figure~\ref{fig8}b) \cite{wu2023device}. The working electrode is coated with a biorecognition element, such as enzymes, antibodies, or nucleic acids, which interacts with the target analyte in the sample. The reference electrode maintains a constant potential, serving as a reference point to measure the potential changes at the working electrode. When the target analyte binds to the biorecognition element on the working electrode's surface, it changes the local charge distribution, resulting in a potential difference. This potential change is directly related to the concentration of the target analyte in the sample. Potentiometric biosensors offer several advantages, including high specificity, label-free detection, and simple instrumentation \cite{ozdemir2013label}. They are particularly suitable for measuring ion concentrations, pH levels, and other analytes directly affecting local charge distribution. 

\textbf{Impedimetric Sensing}: Impedimetric biosensors are a type of electrochemical biosensor that measures the electrical impedance or resistance changes at the sensor's working electrode surface in response to the interaction between a biorecognition element and the target analyte (Figure~\ref{fig8}c). This label-free and real-time detection method is highly sensitive and enables the study of various biomolecular interactions, making it valuable in biosensing applications. The basic setup of an impedimetric biosensor includes a working electrode coated (or functionalized) with a biorecognition element (such as antibodies, enzymes, or DNA probes), a reference electrode, and a counter electrode. When the target analyte (e.g., antigen, enzyme substrate, or complementary DNA strand) binds to the biomolecules, it causes a change in the dielectric properties or the electrical double layer at the electrode surface. When an AC signal is applied to the working electrode, the impedance of the sensor changes due to the binding events between the biorecognition element and the target analyte. These changes in impedance are then measured and correlated with the concentration of the target analyte in the sample.

Impedance-based biosensors can be classified into two main types: capacitive and conductive. Capacitive impedance biosensors rely on changes in the dielectric properties of the interface between the sensing element and the target analyte. When the analyte binds to the immobilized biomolecules, it alters the dielectric constant and thickness of the insulating layer, leading to changes in the electrode's capacitance. These changes are then measured and related to the concentration of the analyte. Conductive impedance biosensors work based on changes in the electrical resistance at the electrode interface. The binding of the analyte to the sensing element causes changes in the electrical properties of the surface layer, leading to variations in resistance. These changes are measured to quantify the analyte concentration.

Impedimetric biosensors offer several advantages, including label-free detection, high sensitivity, and real-time monitoring capabilities. They are particularly suitable for detecting biomolecular interactions, such as antigen-antibody binding, enzyme-substrate reactions, and nucleic acid hybridization. Impedimetric biosensors are versatile and can detect various analytes, including proteins, nucleic acids, and small molecules. Although highly sensitive, Impedance-based biosensors may encounter challenges related to specificity, as they could exhibit cross-reactivity with similar molecules. Careful calibration is essential due to the influence of surface effects on impedance measurements. Additionally, complex sample matrices, such as blood or soil, might interfere with impedance measurements, potentially impacting result accuracy. Addressing these issues is crucial to ensure the reliability and applicability of impedance-based biosensors in various scientific and biomedical applications.

\subsection{Optical Sensing}
Optical sensing utilizes the interaction between light and organic materials to detect and quantify biological or chemical analytes. These sensing platforms employ organic materials, such as organic semiconductors, fluorescent dyes, or organic nanoparticles, integrated into photonic or optoelectronic devices to facilitate sensitive and selective detection of target molecules. The fundamental principle behind optical sensing in organic bioelectronics relies on the optical properties of the organic materials, which can absorb, emit, or scatter light in response to changes in their environment. Within the realm of optical biosensors, various types have been developed, each catering to specific applications and detection requirements \cite{chen2020optical,singh2023optical}. 

\begin{figure}
\centering
\includegraphics[width=\textwidth]{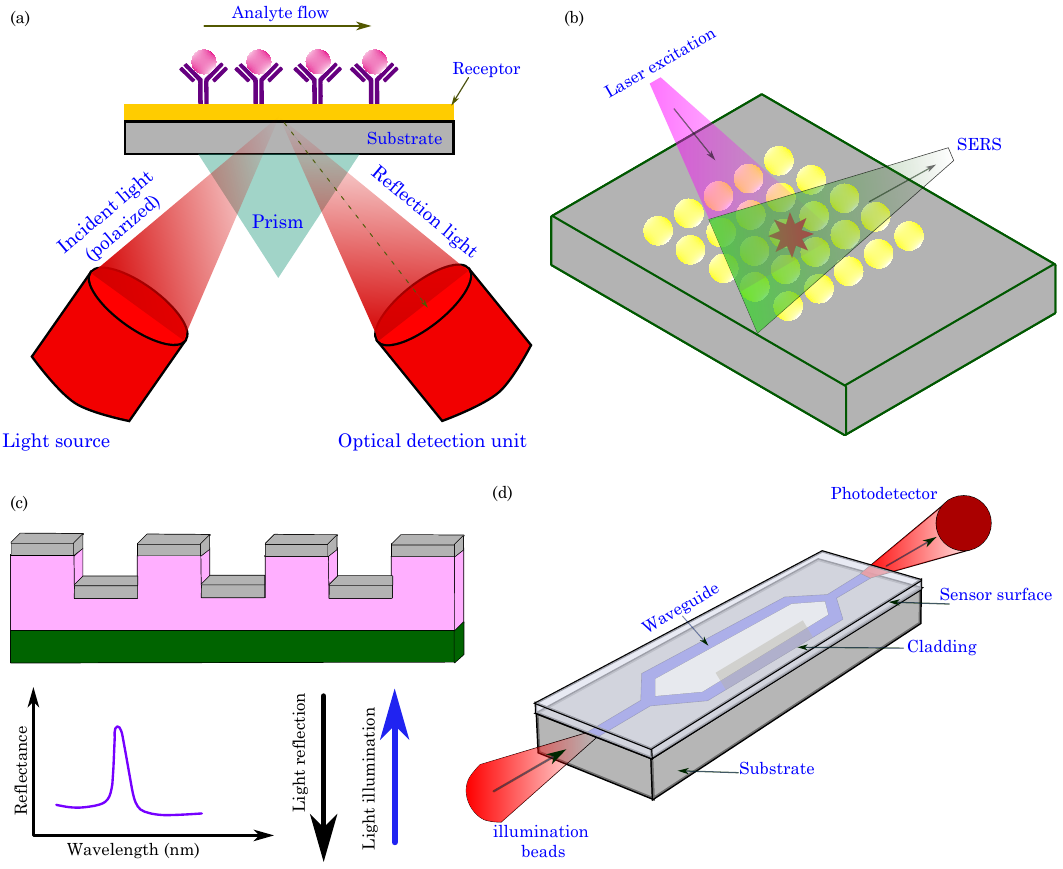}
\caption{Schematics diagrams of optical biosensors. \textbf{(a)} Surface plasmon resonance (SPR) biosensor; \textbf{(b)} Surface-enhanced Raman scattering (SERS) biosensor; \textbf{(c)}  Illustration of the sensing mechanism of a photonic crystal (PC) biosensor. Adapted from Chen et al. \cite{chen2020review}, \emph{Biosensors 2020, 10, 12, 209}, \copyright2020 MDPI; \textbf{(d)} optical waveguide (Mach–Zehnder) interferometer biosensor, adapted with permission from Kozma et al. \cite{kozma2014integrated}, \emph{Biosens. Bioelectron., 2014, 58, 287-307}, \copyright2014 Elsevier B.V.}
\label{fig9}
\end{figure}

Surface plasmon resonance (SPR) biosensors, one of the most well-known optical biosensors, rely on the principle of plasmon resonance, which occurs when light interacts with the collective oscillations of electrons on a metal surface \cite{damborsky2016optical}. Changes in refractive index due to binding events on the sensor surface lead to alterations in the resonance angle, enabling label-free and real-time detection of molecular interactions. Figure~\ref{fig9}(a) shows the schematic of the SPR-based biosensor. SPR biosensors find applications in drug discovery, medical diagnostics, and environmental monitoring \cite{cooper2002optical}.

Surface-enhanced Raman scattering (SERS) biosensors leverage the enhancement of Raman scattering signals when molecules are adsorbed on roughened metal surfaces. Molecules adsorbed on these surfaces generate unique Raman spectra, enabling molecular identification and quantification (Figure~\ref{fig9}b). SERS stands out as an exceptionally sensitive method for identifying low-concentration molecules. It excels in detecting various substances, such as DNA, microRNA, proteins, blood components, and bacteria. Furthermore, it facilitates the detection and characterization of individual cells, aids in bioimaging, and plays a pivotal role in diagnosing various diseases. Its unique ability to offer extensive structural insights into biological analytes adds significant value to the field of analytical science and diagnostics \cite{liang2021carbon}.

Fluorescence is a widely used optical phenomenon for biosensing \cite{borisov2008optical}. In fluorescence-based optical sensing, organic fluorescent dyes or fluorophores are used as the sensing elements. When excited with a specific wavelength of light, these fluorescent molecules absorb energy and become excited to higher energy states. Subsequently, they release this excess energy through fluorescence emission at a longer wavelength. The intensity of the emitted fluorescence signal is directly proportional to the concentration of the target analyte, enabling quantitative detection. Fluorescence-based organic bioelectronic sensors offer high sensitivity and excellent selectivity, making them valuable tools in molecular imaging, cellular assays, DNA sequencing, protein-protein interaction studies, and diagnostic applications. 

Photonic crystal optical biosensors harness the unique properties of photonic crystals to enable sensitive and specific detection of biomolecular interactions \cite{khani2022optical}. These biosensors operate on the principle of modifying the transmission or reflection of light when target molecules bind to the sensor surface. Photonic crystals are engineered materials with periodic structures that create band gaps in the electromagnetic spectrum (Figure~\ref{fig9}c). These band gaps prevent the propagation of certain wavelengths of light, resulting in specific optical properties. When biomolecules bind to the sensor surface, they cause changes in the refractive index or the dielectric environment. This perturbation affects the photonic band gap, leading to light transmission or reflection alterations. These shifts are then used to quantify the presence or concentration of the target analyte.

Interferometric biosensors utilize the interference patterns generated when light waves interact. By measuring changes in phase or intensity, these sensors detect biomolecular interactions. Fabry-Perot interferometers and Mach-Zehnder interferometers (see Figure~\ref{fig9}d) are commonly used in this category. A Fabry-Perot interferometer exploits multiple-beam interference within a resonant optical cavity to precisely measure the wavelengths of light. It consists of two parallel mirrors with a small separation distance, creating a resonant cavity. When light is introduced into the cavity, it reflects repeatedly between these mirrors, leading to constructive and destructive interference between the multiple reflected beams. Constructive interference enhances the intensity of light at specific wavelengths, while destructive interference reduces it at others, producing a pattern of interference fringes. By analyzing these fringes and their variations, Fabry-Perot interferometers can be used to determine the wavelengths of light and facilitate high-resolution spectral analysis. Mach-Zehnder interferometers are typically used in integrated optical biosensors. They consist of two parallel waveguides; one is exposed to the sample, and the other serves as a reference. Biomolecular interactions on the sample waveguide cause changes in optical path length, leading to interference patterns that can be used to quantify the interactions. Interferometric biosensors have applications in medical diagnostics and environmental monitoring.

Optical fiber biosensors employ optical fibers as a core component for detecting and quantifying biological or chemical substances. These sensors are characterized by their capacity to harness light transmission through optical fibers for sensitive and real-time detection. The basic operation typically involves a recognition element, such as antibodies, enzymes, or other bioactive molecules, immobilized on the fiber's surface. When the target analyte binds to this recognition element, it changes the fiber's optical properties, such as light intensity, wavelength, or polarization. These changes are then quantified and correlated to the concentration of the target analyte. These sensors are compact, versatile, immune to electromagnetic interference, and suitable for remote sensing. 

Organic bioelectronic optical sensors offer several advantages, including label-free detection, high sensitivity, rapid response times, and the potential for miniaturization and integration with other electronic components. As organic bioelectronics advances, further research and development of novel organic materials and innovative sensing platforms are expected to drive progress in optical sensing and its applications in various scientific and technological domains.

\subsection{Piezoelectric Sensing}
Piezoelectric biosensing is a powerful and versatile real-time mechanism to detect and quantify biomolecular interactions. This sensing mechanism leverages the piezoelectric effect of certain materials, such as quartz or piezoelectric polymers, to transduce biomolecular binding events into measurable electrical signals. These mass-based biosensors are widely used in biomedical research, diagnostics, and pharmaceutical development due to their label-free, sensitive, and rapid detection capabilities.

The fundamental principle behind piezoelectric biosensing lies in the piezoelectric materials' ability to convert mechanical stress into electrical signals. The biosensing platform typically consists of a piezoelectric transducer, such as a quartz crystal microbalance (QCM) or a piezoelectric polymer-coated cantilever, functionalized with specific biorecognition elements \cite{skladal2016piezoelectric,pohanka2018overview}. These biorecognition elements, such as antibodies, DNA, or enzymes, are carefully immobilized on the surface of the piezoelectric material. When the biosensing platform comes into contact with a biological sample, such as a solution containing biomolecules of interest (e.g., proteins, DNA, or antigens), the biorecognition elements interact selectively with the target biomolecules. This interaction leads to the forming of biomolecular complexes, causing an increase in the mass or stiffness of the layer attached to the piezoelectric material.

\begin{figure}
\centering
\includegraphics[width=\textwidth]{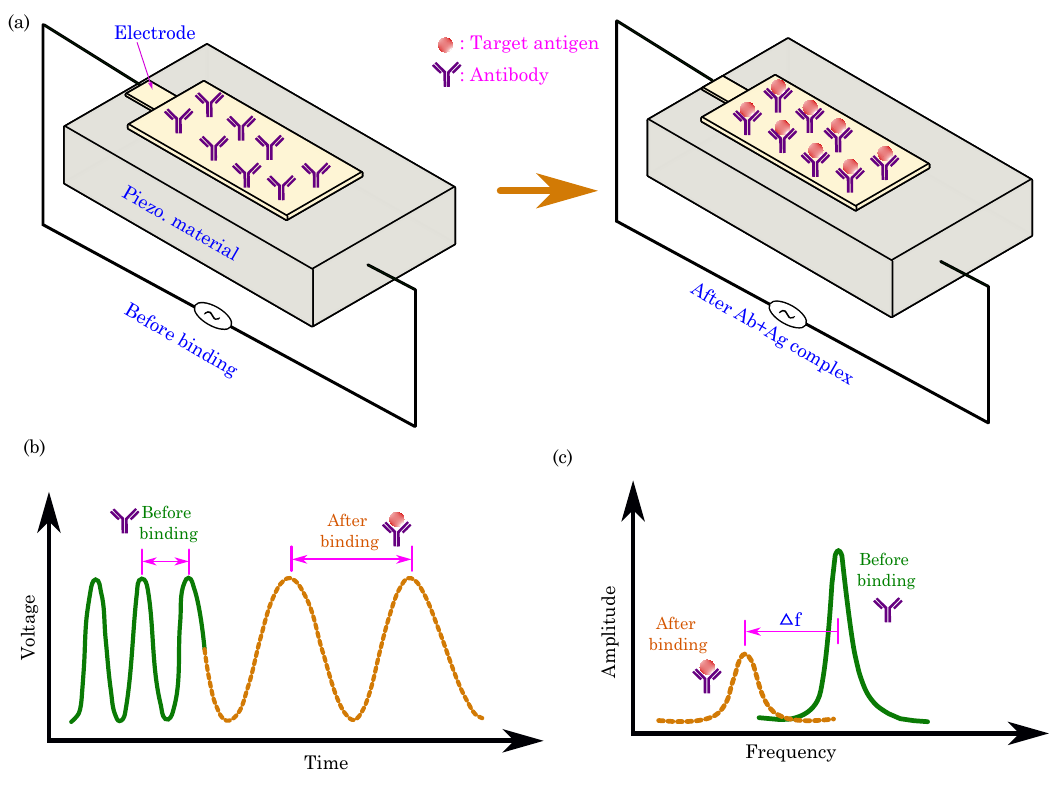}
\caption{\textbf{(a)} Basic concept of target antigen detection mechanism using piezoelectric biosensing, \textbf{(b)} schematics of voltage vs. time, and \textbf{(c)} and amplitude vs. frequency plots during detection.}
\label{fig10}
\end{figure}

 As the biomolecular complexes form, the mechanical stress on the piezoelectric material changes, inducing a shift in the resonant frequency of the piezoelectric transducer \cite{narita2021review}. This frequency shift is directly proportional to the mass or stiffness change on the transducer's surface and is known as the frequency shift or resonance frequency shift. The piezoelectric material converts this mechanical deformation into an electrical signal, generating a characteristic impedance change or charge distribution on the electrode surfaces. Figure~\ref{fig10} shows the basic concept of piezoelectric sensor-based virus detection.

The interaction between the biorecognition elements and target biomolecules can be quantified and analyzed by monitoring the real-time frequency shift or electrical signals. This label-free detection approach directly measures biomolecular interactions without fluorescent or radioactive labels, which can alter biomolecules' behavior and affect the measurements' accuracy. Piezoelectric biosensors offer several advantages in bioanalytical applications, including higher sensitivity, real-time monitoring-label-free sensing, multiplexing-enabling simultaneous detection of multiple target biomolecules in a single experiment, and require low sample volumes-making them suitable for analyzing limited or precious samples.

\section{Biosensing Applications}
\subsection{Medical diagnostics}
Organic bioelectronics has emerged as a promising technology in medical diagnostics, offering unique advantages for non-invasive and point-of-care testing. By leveraging organic materials' electrical and biological properties, organic bioelectronics facilitates the development of sensitive, portable, and cost-effective diagnostic devices \cite{koklu2021organic, bhalla2015introduction,coles2022fluidic}. 

Organic bioelectronic biosensors have opened up new possibilities in disease biomarker detection, enabling the identification of specific biomolecules in biological fluids like blood, saliva, and urine \cite{macchia2020organic,wang2022covalent}. These biosensors can be customized to detect disease-related biomarkers associated with conditions such as cancer, cardiovascular disorders, and infectious diseases, facilitating early diagnosis and timely intervention. In the realm of diagnostics, organic bioelectronics plays a central role in the miniaturization of diagnostic platforms, giving rise to lab-on-a-chip (LOC) devices \cite{parkula2020harnessing,macchia2022handheld}. LOC diagnostics offer rapid and multiplexed testing with minimal sample volume requirements, making them ideal for point-of-care settings and reducing the strain on centralized healthcare facilities. The use of organic bioelectronics extends to electrochemical and electronic immunoassays, providing highly sensitive and specific detection of antigens and antibodies. These assays allow for precise quantification of disease-related molecules, supporting accurate diagnosis and monitoring of disease progression. Nucleic acid analysis is another application of organic bioelectronics, enabling the detection of DNA and RNA sequences associated with genetic disorders and infectious agents \cite{zhang2022overcoming,bai2023electrochemical}. This technology is essential for genetic screening, personalized medicine, and pathogen identification. In medical imaging, organic bioelectronics has shown promise in developing imaging probes and contrast agents, enhancing the resolution and sensitivity of imaging techniques like magnetic resonance imaging (MRI)\cite{fang2015organic,lee2023implantable}. Additionally, organic bioelectronics has contributed to advancing microfluidic systems for cell analysis, enabling cell sorting, counting, and characterizing cellular responses to external stimuli \cite{cheng1998isolation,hsiao2022microfluidic,hsiao2023pedot}. These systems have diverse applications in cancer diagnostics, drug screening, and stem cell research.

Furthermore, the potential for smart drug delivery systems arises from organic bioelectronics, allowing for targeted drug delivery that responds to specific biological signals or conditions, enhancing drug efficiency while minimizing side effects. Also, the portability and affordability of organic bioelectronic devices have made them a viable option for point-of-care diagnostics in resource-limited settings, offering timely and reliable medical testing in underserved regions. Altogether, organic bioelectronics is proving to be a transformative technology in medical and environmental applications, contributing to improved healthcare, diagnostics, and research endeavors. 

\begin{figure}
\centering
\includegraphics[width=0.9\textwidth]{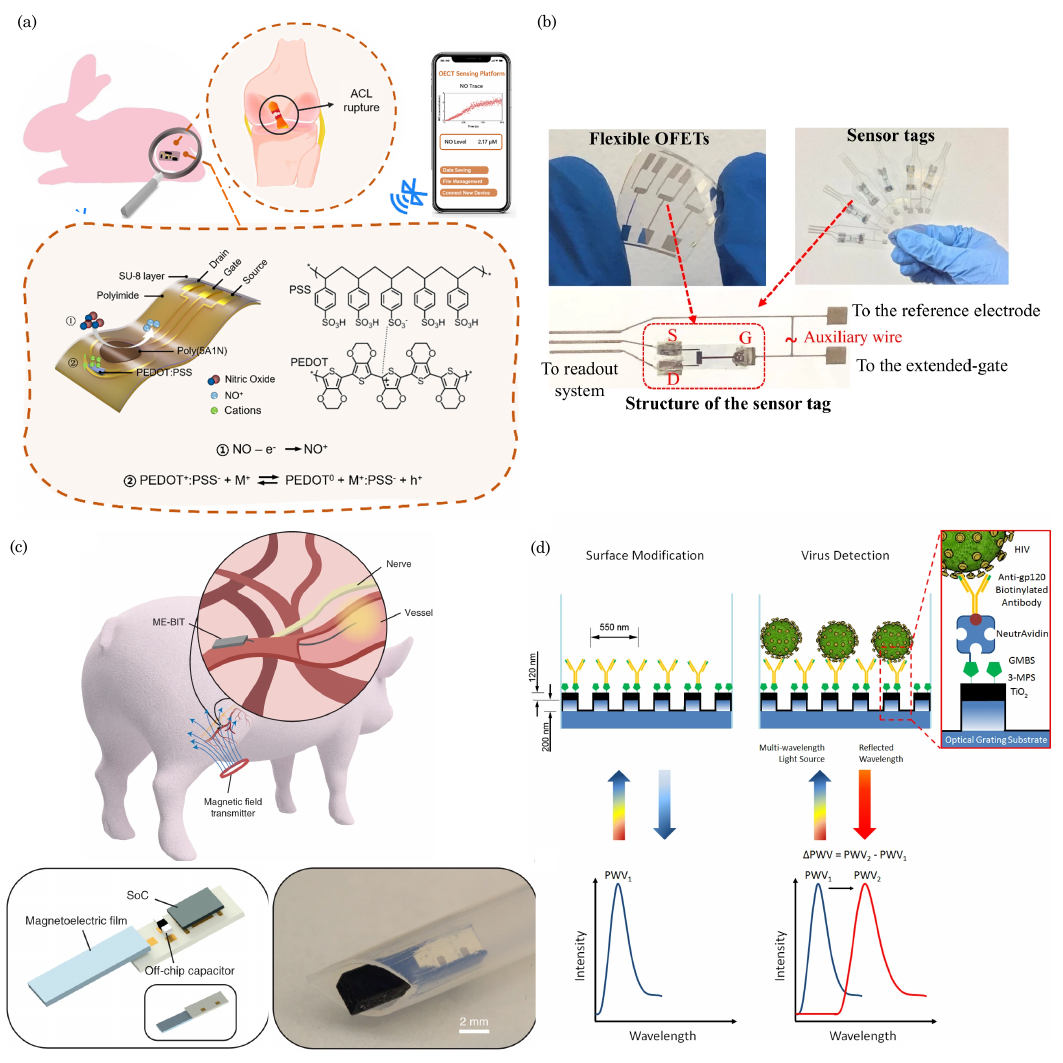}
\caption{\textbf{(a)} Schematic illustration of a flexible OECT Biosensor with Wireless Integration for Real-time NO Detection in an Articular Cavity. The NO sensor features a PEDOT:PSS channel, Au thin film electrodes (source, drain, gate), poly-5A1N selective membrane on the gate, and SU-8 encapsulation exposing specific regions on a PI substrate. NO-induced electrochemical reactions on the gate electrode modulate PEDOT:PSS channel doping, enabling NO sensing via current measurements. Implanted in a New Zealand White rabbit with ACL rupture, the sensor provides real-time NO monitoring, transmitting data to a mobile phone via a Bluetooth-enabled custom wireless module. Deng et al. \cite{deng2022flexible}, \emph{PNAS, 2022, 119, 34, e2208060119}, \copyright2022 the Author(s), licensed under Creative Commons Attribution-NonCommercial-NoDerivatives 4.0 International (CC BY-NC-ND 4.0); \textbf{(b)} Photo images of the fabricated low-voltage OFET miRNA sensor on PEN substrate, and the sensor tags consisting of an encapsulated OFET and contacts for extended-gate sensing electrode and reference electrode. Reprinted from Tang et al. \cite{tang2022solution}, \emph{npj Flex. Electron., 2022, 6,18}, licensed under a Creative Commons Attribution 4.0 International License; \textbf{(c)} Specific illustration of MagnetoElectric-powered Bio ImplanT (ME-BIT) device implanted proximally to a blood vessel deep within tissue and wirelessly powered through a magnetic coil in a pig. A rendering of the implant (\emph{bottom left}) is shown with all the external components, including the system on a chip (SoC), external capacitor, and the ME transducer. Photograph of the fully packaged device inside a 3D-printed capsule resting in a clear sheath (\emph{bottom right}). Reprinted with permission from Chen et al. \cite{chen2022wireless}, \emph{Nat. Biomed. Eng., 2022, 6, 706–716}, licensed under a Creative Commons Attribution 4.0. ; \textbf{(d)} Nanostructured Optical Photonic Crystal Biosensor for HIV Viral Load Measurement. reprinted with permission from Shafiee et al.\cite{shafiee2014nanostructured}, \emph{Sci. Rep., 2014, 4, 4116}, licensed under a Creative Commons Attribution-NonCommercial-NoDerivs 3.0 Unported License.}
\label{fig11}
\end{figure}

Figure~\ref{fig11} displays diverse applications of organic bioelectronics in the field of medical diagnostics. Deng et al. \cite{deng2022flexible} introduced a wireless, flexible, and highly sensitive biosensor employing organic electrochemical transistors (OECTs) for continuous and wireless nitric oxide (NO) detection within biological systems. Their OECT device, depicted in Figure~\ref{fig11}, incorporated a PEDOT:PSS channel, gold (Au) thin film electrodes (source, drain, and gate), a poly-5A1N-coated gate, and electrical contacts on a polyimide (PI) substrate. This sensor was successfully implanted in a rabbit for real-time NO monitoring, with data transmitted wirelessly to a mobile phone via a custom Bluetooth module. Tang et al. \cite{tang2022solution} developed a low-power organic field-effect transistor (OFET)-based biochemical sensor with high transconductance efficiency for label-free miR-21 detection, as seen in Figure~\ref{fig11}(b). Additionally, Chen et al. \cite{chen2022wireless} presented a compact wireless magnetoelectric endovascular neural stimulator illustrated in Figure~\ref{fig11}(c), specifically designed for battery-free implants, enabling stimulation of peripheral nerves that are typically challenging to access via traditional surgical means.

\subsection{Wearable Health Monitors}
Organic bioelectronics has gained considerable traction as a technology for wearable health monitoring systems, offering exceptional versatility and performance. Wearable devices can seamlessly integrate into daily life by leveraging organic materials' unique properties, including flexibility, biocompatibility, and tunable electronics \cite{lin2022flexible,kim2023wearable}. Applying organic bioelectronic sensors allows for the continuous and non-invasive monitoring of vital signs, such as heart rate \cite{nathan2017particle}, blood pressure \cite{kireev2022continuous}, respiration rate \cite{kano2017fast,roy2019self}, body temperature \cite{trung2018stretchable}, pulse \cite{yang2019new}, glucose levels in individuals with diabetes \cite{karpova2019noninvasive}, pH levels \cite{nyein2016wearable}, and the human stress hormone cortisol \cite{parlak2018molecularly}. Also, organic wearable bioelectronics has been widely used for chronic wound biosensing and on-demand therapy administration \cite{wang2022wearable,shirzaei2023stretchable}. 

Furthermore, organic bioelectronics enables the recording of electrocardiogram (ECG) signals for early detection of cardiac abnormalities while monitoring skin conditions, muscle activity during physical activities, sleep patterns, stress levels, and emotions, contributing to comprehensive health assessment \cite{campana2014electrocardiographic,ahmed2020early,yang2023wearable}. These wearable systems can also track environmental factors like air quality and temperature and provide secure biometric authentication for enhanced data security. By combining diverse functionalities, organic bioelectronics empowers individuals to control their health proactively, enabling real-time remote monitoring, personalized drug delivery, and improved overall health management and outcomes \cite{kakria2015real,lee2021standalone}. As research continues, further advancements in organic bioelectronics promise to revolutionize wearable health monitoring technology and its potential impact on healthcare.

\begin{figure}
\centering
\includegraphics[width=0.9\textwidth]{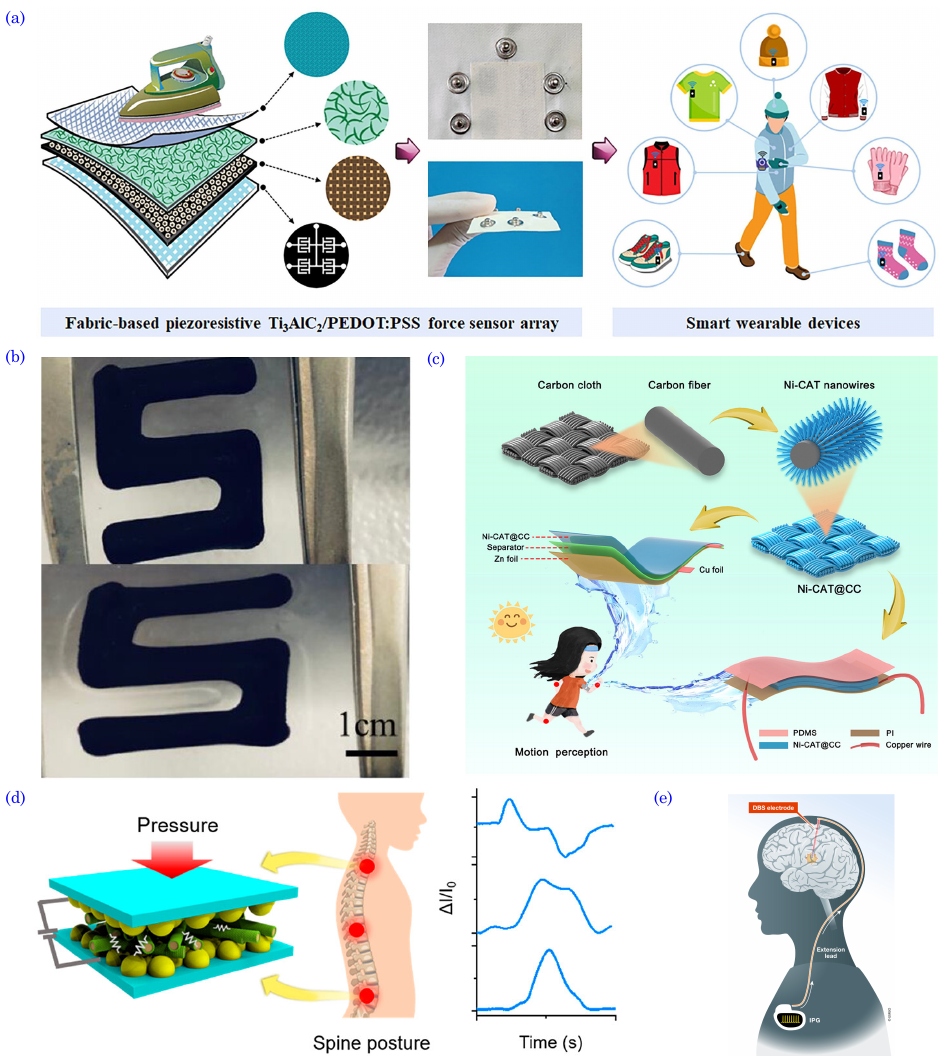}
\caption{Examples of organic bioelectronics-based sensors for wearable health monitoring applications. \textbf{(a)} fabric-based piezoresistive force sensor array based on Ti3AlC2/PEDOT:PSS nanocomposite for wearable E-textile applications. Reprinted with permission from Seesaard and Wongchoosuk \cite{seesaard2023fabric}, \emph{Org. Electron., 2023, 122, 106894}. \copyright2023 Elsevier B.V.; \textbf{(b)} photographs of a stretchable photodiode made of a composite light absorber (P3HT:PCBM:SIS = 1:1:5) on a PDMS substrate before and after being stretched to about 25\% strain. Adapted with permission from Mao et al. \cite{mao2023stretchable} \emph{ACS Appl. Mater. Interfaces 2023, 15, 28, 33797-33808}, \copyright2023 American Chemical Society.; \textbf{(c)} illustration depicting nickel-based metal-organic framework (MOF) nanowires employed as dual-purpose electrodes in wearable pressure sensor technology. Reprinted with permission from Fan et al. \cite{fan2023free}, \emph{iScience 2023, 26, 8, 107397}, \copyright2023 The Author(s); \textbf{(d)} hierarchically microstructure-bioinspired flexible piezoresistive sensor or human–machine interaction and human health monitoring.  The sensor incorporates a hierarchical polyaniline/polyvinylidene fluoride nanofiber (HPPNF) film positioned between two interlocking electrodes featuring a microdome structure. Reprinted with permission from Yang et al. \cite{yang2021hierarchically}, \emph{ACS Nano 2021, 15, 7, 11555–11563}, \copyright2021 American Chemical Society; \textbf{(e)} Schematic diagram for clinical application of deep brain stimulation (DBS) System: The brain electrode delivers therapeutic electrical currents, while the extension lead links it to the neurostimulator (internal pulse generator, IPG), which serves as the implanted power source. Reprinted with permission from Jacobs et al. \cite{jakobs2019cellular} \emph{EMBO Molecular Medicine, 2019, 11, e9575}, \copyright2019 The Author(s), published under the terms of the CC BY 4.0 license.}
\label{fig12}
\end{figure}

Figure~\ref{fig12} exemplifies applications of organic bioelectronics in wearable health monitoring and neuromodulation. Seesaard and Wongchoosuk \cite{seesaard2023fabric} introduced a fabric-based piezoresistive force sensor array composed of a Ti$_3$AlC$_2$/PEDOT:PSS nanocomposite with ultrahigh sensitivity (up to 1.51 N$^-1$) suitable for wearable E-textile applications. In another study, Mao et al. \cite{mao2023stretchable} developed a soft, stretchable photodiode with a composite light absorber and an organic bulk heterojunction within an elastic polymer matrix for reliable cardiovascular variable measurements. The developed  
 photodiode effectively monitors variables such as heart rate variability and oxygen saturation over extended periods. 
Fan et al. \cite{fan2023free} fabricated flexible wearable pressure sensors using free-standing conductive nickel metal-organic framework nanowire arrays on carbon cloth. The developed sensor could monitor human activities, including elbow, knee, and wrist bending, as illustrated in Figure~\ref{fig12}(c). Yang et al. \cite{yang2021hierarchically} designed a flexible piezoresistive sensor with a hierarchical polyaniline/polyvinylidene fluoride nanofiber film for monitoring physiological signals and movement (see Figure~\ref{fig12}(d)). Additionally, organic bioelectronics has found application in deep brain stimulation (DBS) for neuromodulation in movement disorders, such as Parkinson's disease, where they connect brain electrodes to neurostimulators for therapeutic purposes, as depicted in Figure~\ref{fig12}(e).

\subsection{Environmental Monitoring}
Organic bioelectronics has demonstrated significant promise across diverse environmental monitoring applications due to its unique attributes, cost-effectiveness, and compatibility with biological systems. These applications achieve more efficient and sustainable monitoring solutions by leveraging organic electronic devices. Key areas of organic bioelectronics application in environmental monitoring include water quality management and monitoring, enabling real-time detection of various pollutants in water bodies; air quality monitoring to track air pollution levels continuously; and soil health assessment, aiding precision agriculture. 

In water quality management, organic bioelectronics is crucial in detecting and quantifying water pollutants such as heavy metals, organic compounds, and microorganisms \cite{tsopela2016development,ceto2016bioelectronic}. Organic bioelectronic sensors offer high sensitivity and selectivity, enabling real-time water quality monitoring in lakes, rivers, and wastewater treatment facilities \cite{jiang2023integrated,krkljes2023multiparameter}. These sensors can help identify contamination sources, assess the effectiveness of water treatment processes, and ensure compliance with regulatory standards, contributing to preserving water resources and safeguarding aquatic ecosystems.

Similarly, organic bioelectronic sensors can assess essential parameters such as nutrient levels, pH, moisture content, and contaminants in soil quality monitoring \cite{kashyap2021sensing,strand2022printed}. Continuous soil monitoring using these sensors aids in precision agriculture, optimizing fertilizer usage, improving crop yield, and preventing soil degradation. By providing accurate and timely data on soil health, organic bioelectronics supports sustainable land management practices and agriculture waste management and promotes soil conservation \cite{fu2022biorefining}. Additionally, organic bioelectronics is utilized for gas sensing, including greenhouse gases and harmful substances, which is critical for climate change studies and emissions control. For example, Alizadeh and colleagues introduced a molecularly imprinted polymer (MIP)-based electrochemical sensor designed to detect 2,4,6-trinitrotoluene (TNT) in environmental samples such as natural waters and soil \cite{alizadeh2010new}. The sensor operates using electrochemical principles, where the interaction between the imprinted polymer and TNT molecules leads to changes in the sensor's electrical properties. Electrochemical techniques can measure and quantify this interaction, offering a sensitive and reliable means of detecting TNT.

Moreover, integrating organic bioelectronics into wearable devices enables individuals to monitor personal exposure to environmental pollutants and allergens, facilitating informed decisions to minimize exposure risks. These sensors' lightweight and portable nature also makes them ideal for monitoring environmental parameters in remote and challenging-to-access areas, valuable for ecological studies and conservation efforts. The ability to network organic bioelectronic devices creates real-time large-scale environmental monitoring networks, contributing to predictive modeling, early warning systems, and informed environmental management decisions. Moreover, organic bioelectronic biosensors offer rapid and precise detection and quantification of water and soil contaminants, including pesticides and heavy metals, aiding analytical assessments of environmental samples. Applying organic bioelectronics in environmental monitoring demonstrates its potential to enhance environmental sustainability, advance ecological understanding, and drive effective decision-making in various domains.

\begin{figure}
\centering
\includegraphics[width=0.9\textwidth]{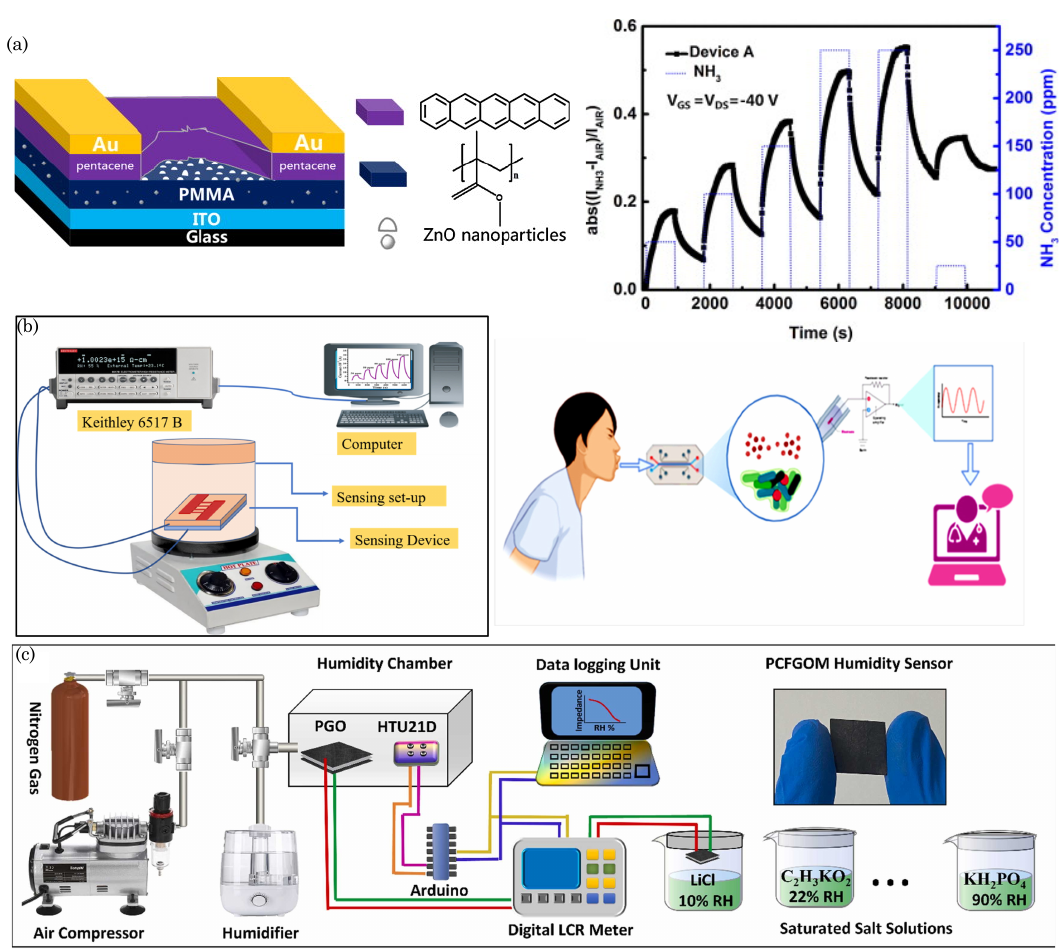}
\caption{ \textbf{(a)} Schematic structure of OFET biosensor for ammonia gas sensing (\emph{left}). In this sensor, poly(methyl methacrylate) (PMMA) blended with zinc oxide (ZnO) nanoparticles is used as a gate dielectric layer. Response curves (\emph{right}) of devices A (ZnO/PMMA hybrid dielectric) exposure to NH$_3$ in higher concentrations (25–250 ppm). Reprinted with permission from Han et al. \cite{han2014performance}, \emph{Sens. Actuators B: Chem., 2014, 203, 9-16}, \copyright2014 Elsevier B.V.; \textbf{(b)} schematic representation of the CuMoO$_4$ nanorods-based acetone sensing measurement setup (left) and non-invasive breathomic-diagnosis of human diabetes and environmental monitoring strategy (right). Reprinted with permission from Mathur et al. \cite{mathur2023cumoo4}, \emph{Environ. Res., 2023, 229,115931}, \copyright2023 Elsevier Inc.; \textbf{(c)} biocompatible paper cellulose fiber graphene oxide matrix-based humidity sensors for human health and environment monitoring. Reprinted with permission from Khan et al. \cite{khan2023biocompatible}, \emph{Sens. Actuators B: Chem.,2023,393,134188}, \copyright2023 Elsevier B.V..}
\label{fig13}
\end{figure}

Figure~\ref{fig13} visually illustrates various organic bioelectronics sensors tailored for environmental monitoring applications. For example, Han et al. \cite{han2014performance} introduced a highly efficient ammonia gas sensor by combining an organic field-effect transistor (OFET) with a ZnO/PMMA hybrid dielectric through a simple blending process. This sensor exhibited remarkable sensitivity across a wide range of NH$_3$ concentrations, from 25 ppm to 250 ppm, as observed in Figure~\ref{fig13}(a) through the time-dependent changes in the drain–source current following multiple NH$_3$ exposure and evacuation cycles. In a separate study, Mathur et al. fabricated CuMoO4 nanorods to create an acetone chemiresistor, enabling non-invasive breath-based diabetes diagnosis and environmental monitoring (depicted in Figure~\ref{fig13}(b)). Khan et al.\cite{khan2023biocompatible} utilized a cellulose fiber and graphene oxide matrix to develop humidity sensors suitable for both environmental humidity monitoring and human respiration detection, as demonstrated in Figure~\ref{fig13}(c).

\subsection{Food Safety and Quality Control}
Organic bioelectronics has emerged as a promising technology for food safety and quality control applications \cite{scognamiglio2014biosensing,rotariu2016electrochemical,curulli2021electrochemical,du2021luminescent}. Its unique properties, including biocompatibility and sensitivity to biological molecules, make it well-suited for detecting contaminants, spoilage, and quality indicators in food products. Key applications include detecting food contaminants like pesticides and pathogens, monitoring food spoilage, assessing food quality indicators, and detecting allergens. Organic bioelectronics allows for real-time monitoring of food production processes and on-site testing, contributing to consistent quality and safety. Additionally, it can be integrated into smart packaging to monitor food quality during storage and transportation. This technology aids in verifying food authenticity, detecting adulteration, and ensuring agricultural production safety by monitoring pesticide residues on crops. Embracing organic bioelectronics in food safety and quality control enhances consumer protection, reduces food waste, and strengthens food safety regulations.

\begin{figure}
\centering
\includegraphics[width=0.9\textwidth]{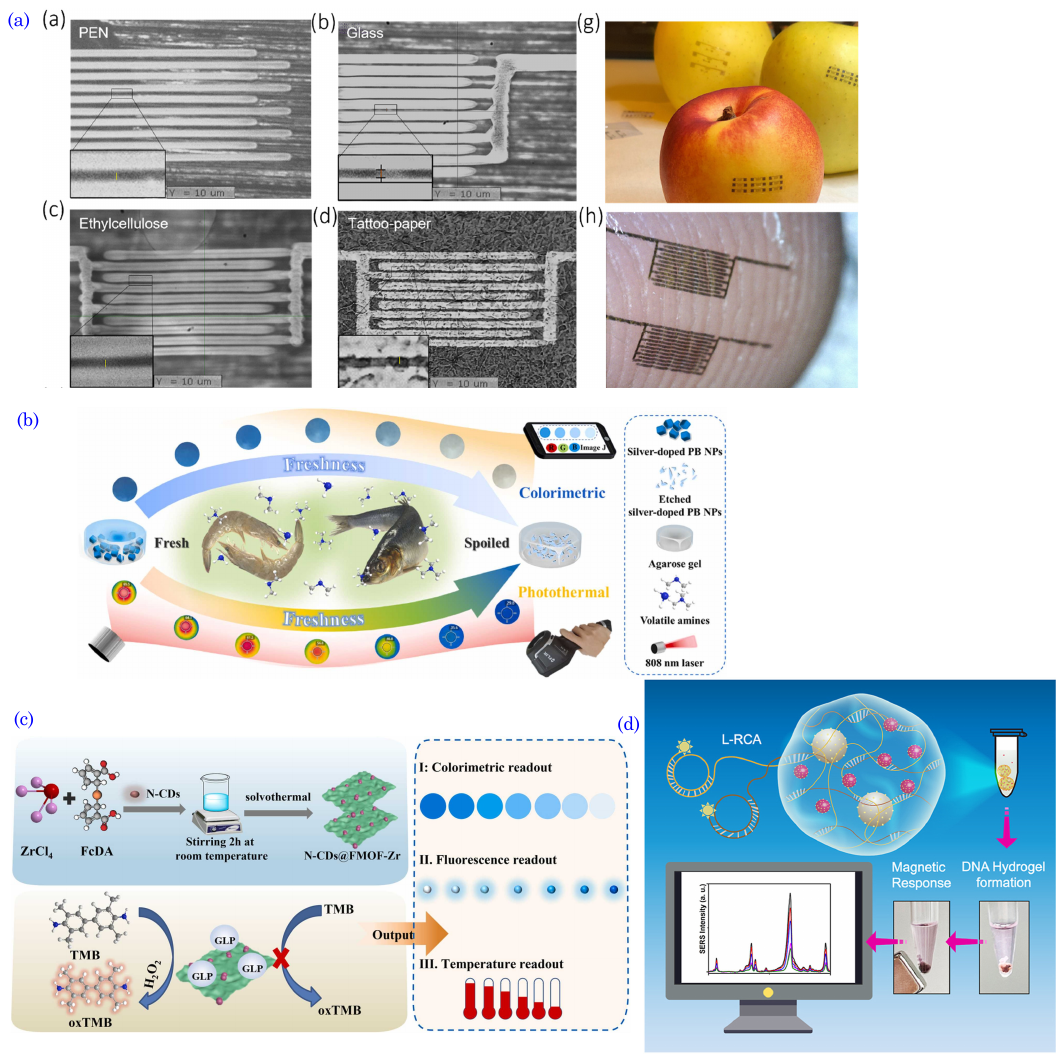}
\caption{\textbf{(a)} Characterization of inkjet-printed gold electrodes on various conventional and edible substrates. \emph{a-d}: depiction of gold interdigitated electrodes inkjet-printed on diverse substrates: poly(ethylene 2,6-naphthalate) (PEN), glass, edible ethyl cellulose biopolymer (food additive E462), and edible tattoo paper. \emph{g,h}: visual representations of gold electrodes transferred onto distinct surfaces: (top) peach, apple, and (bottom) fingertip. Reprinted with permission from Sharova et al. \cite{sharova2023chitosan}, \emph{Nanoscale, 2023,15, 10808-10819}, \copyright2023 The Royal Society of Chemistry; \textbf{(b)} schematic representation of colorimetric and photothermal assessment of shrimp and fish freshness utilizing a portable silver-doped Prussian blue nanoparticles (SPB NPs) hydrogel, facilitated by a smartphone and handheld thermal imager. Reprinted with permission from Ding et al. \cite{ding2022portable}, \emph{Sens. Actuators B: Chem., 2022, 363, 131811}, \copyright2022 Elsevier B.V.; \textbf{(c)} carbon dots anchoring ferrocene metal-organic framework nanosheet for multi-mode glyphosate (e.g., herbicide) sensing. Reprinted with permission from Luo et al. \cite{luo2023versatile}, \emph{J. Hazard. Mater., 2023, 443, 130277}, \copyright2022 Elsevier B.V.; \textbf{(d)} schematic illustration of SERS aptasensor based on DNA hydrogel fishing network for ultrasensitive detection of antibacterial kanamycin (KANA). Reprinted with permission from Chen et al. \cite{chen2022fishing}, \emph{Biosens. Bioelectron., 2022, 207, 114187}, \copyright2022 Elsevier B.V.}
\label{fig14}
\end{figure}

Figure~\ref{fig14} illustrates the diverse applications of biosensors in food safety and quality control. Sharova et al. \cite{sharova2023chitosan} introduced a low-voltage edible electronic circuit, serving as an invaluable testbed for exploring non-toxic printable semiconductors within the domains of edible and bioelectronic technologies. Their work, presented in Figure~\ref{fig14}(a), showcased successful inkjet printing of water-based gold ink on both traditional and edible substrates, achieving exceptional precision with critical lateral features as small as 10 $\mu$m. Furthermore, they demonstrated the fabrication of chitosan-gated complementary n- and p-type transistors and logic circuits, including inverting logic gates, all operating at low voltages (<1 V) on flexible edible ethyl cellulose substrates. These devices exhibited promising electronic performance characteristics, such as high mobility–capacitance product, impressive on–off current ratios, operational stability in ambient air, and a shelf life of up to one month. These devices' compact, flexible nature allows for seamless integration into edible carriers, such as pharmaceutical capsules. In a separate study, Ding et al. \cite{ding2022portable} introduced a hydrogel containing silver-doped Prussian blue nanoparticles (SPB NPs) for the detection of trimethylamine (TMA) and the real-time monitoring of shrimp and fish freshness, as depicted in Figure~\ref{fig14}(b). Additionally, Luo et al. \cite{luo2023versatile} explored using carbon dots anchored to ferrocene metal-organic framework nanosheets for the multi-mode sensing of glyphosate, a herbicide. In another application, Chen et al. \cite{chen2022fishing} employed a DNA hydrogel fishing network for the ultrasensitive detection of the antibacterial agent kanamycin. These diverse applications underscore biosensors' remarkable versatility and potential in enhancing food safety and quality control.

\section{Challenges and Future Perspectives}

\subsection{Stability and Longevity}
Stability and longevity are paramount considerations in applying organic electronics in biosensing, given their unique properties, such as flexibility and biocompatibility \cite{li2021implantable}. However, certain challenges contribute to potential degradation and performance fluctuations over time. First, organic materials are susceptible to environmental factors like moisture, oxygen, and temperature variations, leading to material degradation and subsequent changes in electrical properties, diminishing sensor performance \cite{shim2021physically}. Second, ensuring long-term biocompatibility when these devices interact with biological samples is critical to avoiding adverse reactions and preserving reliable sensing capabilities \cite{feron2018organic}. Third, the stability of the interface between the organic material and biomolecules significantly impacts biosensor performance, with changes from material degradation or biofouling affecting sensitivity and selectivity \cite{song2021materials}. In wearable or implantable devices, the organic materials must endure mechanical stress without functional compromise, and mechanical strain may cause cracks or delamination, jeopardizing stability and longevity \cite{root2017mechanical}. Additionally, variations in performance over time due to charge trapping, ion migration, and relaxation processes can lead to sensor response drift, hampering accuracy. Sensitivity to chemicals and solvents can also affect stability and performance, a critical concern in real-world applications where chemical exposure is expected.

Moreover, organic materials may experience photochemical degradation when exposed to light, especially UV radiation, impacting their electrical properties and sensor performance \cite{yousif2013photodegradation}. Finally, achieving manufacturing consistency and uniformity in organic electronic devices presents challenges, as variations in fabrication processes may lead to device-to-device performance differences, influencing reproducibility and reliability. Addressing these stability and longevity concerns is essential for enhancing organic electronics' long-term viability and effectiveness in biosensing applications.

Several strategies can be employed to address the stability and longevity issues associated with organic electronics in biosensing. These include careful material selection, implementing encapsulation techniques and barrier layers to protect the devices from environmental factors, optimizing device design for mechanical robustness, and performing rigorous testing and validation under relevant environmental conditions. Additionally, surface modifications and integrated control systems can enhance organic biosensors' stability and operational performance. Overall, addressing these challenges will pave the way for the successful integration of organic electronics into cutting-edge biosensing technologies, enabling advancements in medical diagnostics, environmental monitoring, and other critical applications.

\subsection{Biocompatibility, Biofouling, and Cross-sensitivity}
Biocompatibility and biofouling pose significant challenges when utilizing organic electronics in biosensing applications. Despite the advantages of organic materials, such as flexibility and tunable properties, ensuring their compatibility with biological systems and mitigating the impact of biofouling is critical for reliable and long-term biosensor performance. In the realm of biocompatibility challenges, implantable biosensors necessitate favorable interactions between organic electronic materials and surrounding tissues to avoid inflammation or immune responses that may compromise the biosensor's functionality and lifespan. Issues like cytotoxicity and impaired cell adhesion when in contact with biological fluids can disrupt stable biomolecule interactions, leading to unreliable measurements. Additionally, an inflammatory response triggered by organic materials could result in encapsulation or scarring around the biosensor, hindering target analyte diffusion and affecting sensor sensitivity \cite{gray2018implantable}.

Moreover, leaching specific molecules from organic materials into the biological environment may compromise the biosensor's accuracy and specificity. On the other hand, biofouling challenges encompass the non-specific binding of biomolecules, proteins, or cells to the biosensor surface, generating unwanted signals and reducing sensitivity \cite{patel2023multifunctional}. Accumulation of biofilm or organic material on the sensor surface can alter the electrical properties of the organic material, leading to a decline in sensor performance over time. Moreover, biofouling can hinder the diffusion of target analytes to the sensing elements, causing delayed or inaccurate readings and impacting the biosensor's response time. Tackling these biocompatibility and biofouling challenges requires careful material selection, surface modifications, and continuous research and development of innovative strategies to ensure the successful integration of organic electronics in biosensing applications.

Furthermore, cross-sensitivity within the domain of organic bioelectronics encompasses a significant challenge whereby sensors and devices, originally engineered to discern and respond to specific target analytes, also manifest responses to unintended analytes, thereby introducing ambiguity and inaccuracies into the device's output. This pervasive issue permeates throughout the sensor and biosensor realm, including the specialized domain of organic bioelectronics. The implications of cross-sensitivity are noteworthy, encompassing potential distortions or falsifications of data, thereby diminishing the overall precision and reliability of the sensor. Several intricate factors contribute to cross-sensitivity in the context of organic bioelectronics. Firstly, material interactions stemming from the inherent properties of organic materials utilized in bioelectronic devices can predispose them to interactions with multiple analytes, exemplified by conducting polymers that may exhibit sensitivity to variances in pH, humidity, or temperature, potentially fostering cross-sensitivity unless these issues are meticulously mitigated. Secondly, the propensity for analytes with resembling properties to induce overlapping sensor responses poses a significant challenge. For instance, two distinct gases may evoke analogous alterations in electrical conductivity, thus complicating their differentiation. Thirdly, the adsorption characteristics of the sensor's surface may occasion unforeseen interactions with analytes, particularly in sensors reliant on specific binding events, such as antibody-antigen interactions. This can give rise to cross-reactivity when analytes bearing similar structural or property traits adhere to the sensor's surface. Lastly, environmental variables, including shifts in temperature, humidity, or interference from electromagnetic fields, may influence the sensor's response, potentially culminating in unwanted noise or disruptions that aggravate cross-sensitivity concerns. Cross-sensitivity challenges require meticulous consideration when designing, implementing, and employing organic bioelectronic devices, particularly in mission-critical applications such as medical diagnostics and environmental monitoring, where precision and fidelity are indispensable.

Researchers and engineers have explored various strategies to address the biocompatibility and biofouling challenges associated with organic electronics in biosensing \cite{long2019effective,chen2021antifouling}. Surface engineering techniques, such as functionalization with biocompatible coatings or polymers, enhance the biocompatibility of organic materials and reduce non-specific binding \cite{kumar2011biocompatible,alkhoury2020study}. Implementing biocompatible encapsulation materials or membranes isolates the organic electronics from direct contact with biological fluids, minimizing adverse tissue interactions. Coating the sensor surface with antifouling agents prevents the adhesion of biomolecules and reduces the impact of biofouling on sensor performance. Rigorous in vitro and in vivo biocompatibility testing is crucial to identify potential cytotoxicity or inflammatory responses early in development. Employing regeneration methods, such as chemical or enzymatic cleaning, helps restore sensor functionality and combat the effects of biofouling. Continual research and development of new organic materials with improved biocompatibility and resistance to biofouling are essential to advance organic electronics for biosensing applications. 
Additionally, addressing cross-sensitivity in organic bioelectronics necessitates a multi-faceted approach, spanning judicious material selection, intelligent surface functionalization strategies, advanced data processing techniques, and rigorous calibration measures to rectify inaccuracies arising from environmental factors. This multi-pronged strategy not only underscores the complexity of addressing cross-sensitivity in organic bioelectronics but also highlights the necessity for a holistic and integrated approach, where material science, surface engineering, advanced data analysis, and robust calibration regimes converge to mitigate the challenges posed by cross-sensitivity, ultimately contributing to the enhanced accuracy and reliability of organic bioelectronic devices.
 
By addressing these challenges, researchers can enhance the reliability and longevity of organic electronic biosensors, paving the way for their successful integration in a wide range of biosensing applications, from medical diagnostics to environmental monitoring and beyond.

\subsection{Manufacturing Scalability}
Manufacturing scalability poses a critical challenge in utilizing organic electronics for biosensing applications despite the advantages of flexibility and cost-effectiveness offered by organic materials. The endeavor to achieve large-scale and reproducible manufacturing encounters several obstacles. These include maintaining material consistency to ensure uniform sensor characteristics and reliable performance, tackling challenges in scaling up deposition techniques like inkjet printing and spin-coating while preserving sensor integrity, and addressing the complexities of device integration with multiple functional layers. Moreover, managing yield and reproducibility risks, achieving cost-effectiveness, and ensuring stability and reliability in large-scale production are paramount. Robust quality control measures are indispensable for the early identification and resolution of manufacturing issues, encompassing material testing, sensor characterization, and performance validation. A reliable supply chain for high-quality organic materials is also crucial for sustained sensor performance and product reliability in the realm of organic electronics biosensing.

To address the manufacturing scalability challenges related to organic electronics in biosensing, researchers and industry stakeholders are exploring various approaches. Developing innovative and scalable manufacturing techniques, such as lithography, roll-to-roll printing, and spray coating, can improve production efficiency and material utilization \cite{fruncillo2021lithographic,bobrowski2020scalable}. Establishing standardized protocols and optimizing manufacturing processes can enhance yield, reproducibility, and material consistency. Integrating real-time quality control measures during manufacturing can detect deviations and ensure uniform sensor performance. Conducting rigorous, long-term stability testing under various environmental conditions is crucial to assessing sensor performance and reliability over extended periods. The widespread adoption of organic electronics in biosensing applications can be realized by tackling these obstacles, paving the way for developing cost-effective, high-performance biosensors capable of transforming healthcare, environmental monitoring, and other critical domains.

\subsection{Integration and Miniaturization}
Incorporating organic bioelectronics into biosensing devices poses significant challenges in integration and miniaturization. Although organic materials offer unique advantages like flexibility and biocompatibility, achieving seamless integration into compact and multifunctional biosensors requires overcoming various obstacles. Key issues encompass multifunctional integration to create advanced biosensors capable of detecting multiple analytes and coordinating interactions between organic electronic components. Optimizing the sensor-substrate interface when integrating onto diverse substrates is essential to avoid performance degradation \cite{xu2019design}. Power supply and energy efficiency become crucial in miniaturized biosensors operating on limited power sources \cite{xu2020skin}. Maintaining high sensing performance and signal-to-noise ratio in shrinking biosensors is challenging due to signal interference and noise \cite{neshani2023highly}. Precision in fabrication processes and high yield rates are crucial to achieving accurate dimensions and meeting demand while reducing production costs. Efficient data communication and onboard data processing are vital for real-time data transmission in miniaturized biosensors. Ensuring stability, longevity, enhanced biocompatibility, and addressing biofouling challenges are critical to maintaining reliable sensor performance over time in downsized organic bioelectronic components.

Researchers and engineers have employed various strategies to address integration and miniaturization challenges related to organic bioelectronics in biosensing. State-of-the-art microfabrication techniques enable precise control over sensor dimensions and facilitate multi-component integration. Selecting suitable materials and optimizing sensor-substrate interfaces ensures compatibility and mechanical stability in miniaturized biosensors \cite{song2021materials}. Designing low-power circuits and exploring energy-efficient strategies (e.g., self-powered sensors) extend miniaturized biosensors' battery life and autonomy \cite{lundager2016low,song2021self}. Implementing noise reduction techniques and signal amplification methods enhance the signal-to-noise ratio in miniaturized biosensors \cite{dweiri2015ultra,ramesh2022nanotechnology,wang2022electrochemical}. Utilizing automated manufacturing processes ensures reproducibility and precision, while robust quality control measures identify defects early in production. System-on-chip integration enables onboard data processing, reducing the need for external data handling devices \cite{fischer2015integrating,zhao2021responsive,stuart2022wearable}. Applying biocompatible coatings to miniaturized biosensors improves biocompatibility and reduces biofouling \cite{ruiz2018delayed,xu2020anti,chan2022combinatorial}. Effectively addressing these challenges empowers organic bioelectronics to pave the way for highly compact and versatile biosensors with applications ranging from wearable health monitoring to point-of-care diagnostics, thereby advancing healthcare and biosensing capabilities.

\subsection{Data Security and Privacy}
Data security and privacy are crucial concerns in the context of using organic electronics in biosensing applications \cite{camara2015security,mclamore2021feast,ibrahim2022futuristic}. With sensitive biological and health-related data being collected by these devices, maintaining the confidentiality and integrity of this information becomes paramount. Key issues include securing data transmission through robust encryption protocols, authenticating the biosensing device and its generated data to prevent tampering and unauthorized access, and ensuring secure data storage with strong encryption and access control measures. It is also essential to implement secure communication protocols between the biosensor and external devices or servers, anonymize and de-identify collected data to protect individual privacy, and guard against cyberattacks like malware and ransomware \cite{sathya2017secured,garg2020bakmp}. Compliance with data protection regulations like GDPR and HIPAA is necessary, as is user awareness and education about data security best practices. A well-defined data breach response plan and proper data erasure procedures at the end of a device's life cycle are additional measures to mitigate risks and ensure the ethical use of biosensor data. As organic electronics advance in biosensing, a comprehensive approach to data protection is essential to foster trust and safeguard sensitive information.

\subsection{Future perspectives of organic bioelctronics}

Recent times have witnessed a revolutionary transformation in biosensor technology, achieved through synergistic integration with cutting-edge technologies such as smartphones, 3D printing, artificial intelligence, and the Internet of Things (IoT) \cite{zhang2022intelligent}. This convergence has led to unprecedented advancements and opportunities in biosensors. By leveraging the capabilities of these emerging technologies, biosensors have become more accessible, versatile, and efficient than ever before. Smartphones now serve as portable and user-friendly interfaces for real-time data collection and analysis, making biosensing widely accessible. 3D printing has enabled the rapid prototyping and customization of biosensors, allowing tailored designs to meet specific application requirements. Artificial intelligence has empowered biosensors with advanced data processing and pattern recognition capabilities, enhancing accuracy and enabling predictive analytics. The IoT has facilitated seamless connectivity and remote monitoring of biosensors, enabling real-time data transmission and applications in remote and distributed environments. This amalgamation has opened new horizons in healthcare, environmental monitoring, food safety, and beyond, reshaping the future of biosensor applications.

Moreover, the trajectory of organic bioelectronics in intelligent biosensing strategies holds immense promise due to rapid technological progress and interdisciplinary collaborations. This trajectory envisions multiple transformative directions that underline the potential evolution of intelligent biosensing using organic bioelectronics. These include the development of smart biosensing platforms that can autonomously make decisions and incorporate artificial intelligence algorithms for real-time analyte detection and quantification \cite{manickam2022artificial}. Additionally, there's a growing focus on sensors that can self-calibrate using internal or external reference signals to enhance accuracy and reliability \cite{bresnahan2021autonomous,li2021closed,kang2022development}. Integrating data from different sensors employing diverse sensing modalities promises a more comprehensive understanding of sample composition. The concept of dynamic sampling, where sensors adapt their sampling rates based on detected analyte shifts, could optimize energy usage while ensuring timely detection. Furthermore, realizing interconnected sensing networks, predictive analytics, human-machine interfaces, personalized medical interventions, energy-efficient designs, and remote monitoring through telehealth services showcases the broad scope of organic bioelectronics' role in revolutionizing intelligent biosensing \cite{vaghasiya2023wearable}. 

Furthermore, the advancement of sustainable organic bioelectronic sensors holds significant promise, propelled by progress in materials science, biotechnology, and a growing environmental consciousness. These sensors are increasingly capable of utilizing biodegradable and environmentally friendly materials, minimizing their ecological footprint \cite{granelli2022high}. The integration of energy-harvesting technologies further lessens their dependence on traditional batteries by tapping into renewable sources like solar energy or vibrations \cite{bandodkar2017soft,zou2021recent}. The potential for mass production of flexible and printable organic electronics opens doors to versatile applications, including healthcare and environmental monitoring. Moreover, affordable, sustainable bioelectronic sensors are pivotal in addressing global health challenges facilitating remote disease monitoring in resource-limited regions.

\section{Conclusions}
Organic electronics in biosensing represent a promising and dynamic frontier with far-reaching implications for medical and environmental applications. This exciting convergence of organic materials and bioelectronics has unlocked new opportunities for precise, sensitive, and real-time detection of biomolecules and chemical species, transforming the landscape of medical diagnostics and environmental monitoring.

The unique properties of organic materials, such as biocompatibility, flexibility, and tunability, have paved the way for developing innovative biosensing devices with diverse applications. From implantable biosensors for continuous health monitoring to wearable devices enabling personalized diagnostics, organic bioelectronics offers groundbreaking solutions that bridge the gap between traditional sensing technologies and cutting-edge medical practices. In medical diagnostics, organic bioelectronic sensors offer the potential to revolutionize disease detection and management. These sensors' label-free and real-time monitoring capabilities enable rapid and accurate analysis of biomarkers, facilitating early disease diagnosis and tailored treatment plans. Moreover, integrating organic bioelectronics into wearable health monitoring systems empowers individuals to actively participate in their healthcare, promoting proactive and personalized health management. Beyond medical applications, the versatility of organic bioelectronics finds significant relevance in environmental monitoring. From detecting pollutants and toxins to monitoring changes in environmental parameters, organic bioelectronic sensors contribute to sustainable environmental management and conservation efforts. These sensors offer the potential for rapid and efficient detection of environmental threats, enabling timely interventions and preserving ecological balance. However, as with any emerging technology, organic electronics in biosensing face challenges that warrant attention. Issues related to biocompatibility, stability, scalability, and manufacturing consistency must be addressed to ensure these biosensing platforms' reliability and long-term performance.

In conclusion, integrating organic electronics in biosensing is promising for medical and environmental applications. With ongoing research and collaborative efforts between scientists, engineers, and industry stakeholders, organic bioelectronics is poised to drive transformative advancements in healthcare and environmental sustainability. By harnessing the potential of organic materials and innovative sensing mechanisms, this frontier of biosensing promises to improve human health, protect the environment, and shape a more sustainable and technologically advanced future.

\bibliographystyle{unsrt}  
\bibliography{manuscript}

\end{document}